\documentclass[10pt,journal]{IEEEtran}

\usepackage{amssymb}
\usepackage{amsmath}
\usepackage{cite}
\usepackage{url}
\usepackage{xcolor}
\usepackage{cite,graphicx,amsmath,amssymb}
\usepackage{fancyhdr}
\usepackage{mdwmath}
\usepackage{mdwtab}
\usepackage{caption}
\usepackage{amsthm}
\usepackage{setspace}
\usepackage{hyperref}
\usepackage{algorithm}
\usepackage{algorithmic}
\usepackage{multirow}
\usepackage{makecell}
\usepackage{mathtools}
\usepackage{subcaption}
\usepackage{bm}
\usepackage{tikz}
\usepackage{xurl}
\usepackage{lipsum}
\usepackage{gensymb}
\usepackage{balance}
\usepackage{threeparttable}
\usepackage{makecell}
\usepackage{diagbox}
\usepackage{stfloats}

\hypersetup{colorlinks=true,
linkcolor=blue,
citecolor=blue,      
urlcolor=black,
}

\newtheorem{remark}{Remark}
\newtheorem{theorem}{Theorem}

\newtheorem{lemma}{Lemma}

\newtheorem{corollary}{Corollary}

\allowdisplaybreaks
\title{Multiport Network Modeling and Optimization for Reconfigurable Pinching-Antenna Systems}

\author{
        Zhaolin Wang, Jiaqi Xu, Chongjun Ouyang, Xidong Mu, and Yuanwei Liu \\
\vspace{0.4cm}
\emph{(Invited Paper)
}
\thanks{\emph{(Corresponding author: Yuanwei Liu)}}
\thanks{Zhaolin Wang and Yuanwei Liu are with the Department of Electrical and Electronic Engineering, The University of Hong Kong, Hong Kong (e-mail: \{zhaolin.wang, yuanwei\}@hku.hk).}
\thanks{Jiaqi Xu is with the Department of Electrical Engineering and Computer Science, University of California at Irvine, Irvine, CA 92697 USA (e-mail: xu.jiaqi@uci.edu).}
\thanks{Chongjun Ouyang is with the School of Electronic Engineering and Computer Science, Queen Mary University of London, London E1 4NS, U.K. (e-mail: c.ouyang@qmul.ac.uk).}
\thanks{Xidong Mu is with the Centre for Wireless Innovation (CWI), Queen’s University Belfast, Belfast, BT3 9DT, U.K. (e-mail: x.mu@qub.ac.uk).}
}

\begin{document}

\maketitle

\begin{abstract}

    A reconfigurable pinching-antenna system (PASS) is presented, endowing pinching antennas (PAs) with both amplitude- and phase-controllable radiation beyond conventional implementations. To characterize this feature, a general and physically consistent model is established for PASS via multiport network theory. Within this model, the fundamental constraint of ideal reconfigurability of PAs is identified, allowing the full control of signal amplitudes and phases. A practical directional-coupler (DC)-based PA model is then proposed, enabling both amplitude-only control and amplitude-constrained phase control. Beamforming optimization is investigated for both ideal and practical cases: an optimal solution is obtained for ideal PAs, whereas a high-quality iterative algorithm is developed for DC-based PAs. Numerical results suggest that in single-user scenarios: (i) with optimized PA positions, performance gains arise primarily from amplitude reconfigurability and DC-based PAs approach ideal performance, and (ii) with fixed PA positions, both amplitude and phase reconfigurability are critical and DC-based PAs incur non-negligible loss.
\end{abstract}

\section{Introduction} \label{sec:intro}

The relentless demand for ubiquitous, high-capacity wireless connectivity is a defining challenge of our time, driving the evolution towards sixth-generation (6G) networks and the hyper-connected Internet of Things \cite{saad2019vision}. However, the performance of all wireless systems is fundamentally constrained by the physics of wave propagation, namely severe path loss and signal blockage from physical obstacles. While advanced techniques such as massive multi-input multi-output (MIMO) \cite{larsson2014massive} have greatly enhanced spectral efficiency, they primarily compensate for channel impairments rather than eliminate them, leaving the system vulnerable to the large-scale fading effects that dominate in complex environments. This reliance on fixed-position antennas represents a critical bottleneck, limiting reliability and constraining the design of future intelligent and adaptive communication systems. Recently, several techniques have introduced position flexibility of antennas, including fluid antennas \cite{wong2020fluid,new2024tutorial}, movable antennas \cite{zhu2023movable,ning2025movable}, and flexible intelligent metasurfaces \cite{bai2022dynamically,an2025flexible}. However, these approaches still adhere to conventional radio-frequency (RF) front-end architectures and typically permit only wavelength-scale repositioning of radiating elements, which does not fundamentally resolve the aforementioned challenges.

Pinching-antenna system (PASS) is a transformative antenna architecture that shifts the paradigm from channel compensation to direct channel control, which was first introduced by NTT DOCOMO in 2021 together with a working prototype \cite{suzuki2022pinching}. PASS abandons the conventional RF front-end architecture, but is based on low-loss dielectric waveguides that act as signal conduits. By applying small, movable dielectric particles, which are referred to as the “pinching antennas (PAs)”, at arbitrary locations along the waveguide, a set of precisely positioned, on-demand radiation points can be created. This mechanism allows the antenna system to dynamically reconfigure its physical structure, creating strong, line-of-sight (LoS) links that bypass obstructions and minimise free-space propagation distance. By physically positioning the PAs in close proximity to user devices, PASS establishes robust communication channels in environments where conventional systems would fail. Furthermore, this architecture enables a novel beamforming technique. The phase of the signal radiated from each PA can be adjusted by precisely controlling its position along the waveguide, enabling a new beamforming techniques referred to as “pinching beamforming”~\cite{liu2025pinching}.

Owing to these favorable properties, PASS has attracted growing attention and has been extensively investigated from multiple perspectives \cite{ding2025flexible, wang2025modeling, tegos2025minimum, xiao2025frequency, chen2025dynamic, ouyang2025array, jiang2025pinching, papanikolaou2025resolving, xu2025pinching, xu2025joint}. Nonetheless, several critical challenges remain open. First, there is still no unified and physically consistent modeling framework for PASS. Most existing studies establish signal models by analogy with conventional electrical antennas, which overlook the distinct operating principles of PASS. Although physics-based models grounded in coupled-mode theory (CMT) have been developed in \cite{wang2025modeling} and \cite{xu2025pinching}, they suffer from important limitations. In particular, the CMT-based model is effective for downlink characterization but becomes mathematically intractable for uplink analysis, and it fails to handle the practical hardware impairments such as impedance mismatch or fabrication imperfections. Second, the majority of existing works assume non-reconfigurable PAs, so pinching beamforming is achieved only through PA repositioning, which constrains beamforming flexibility. While amplitude-reconfigurable PAs have recently been introduced \cite{xu2025pinching, xu2025joint}, the physical foundations of amplitude control and the feasibility of achieving phase reconfigurability remain largely unexplored.

\begin{figure*}[t!]
  \centering
  \includegraphics[width=0.8\textwidth]{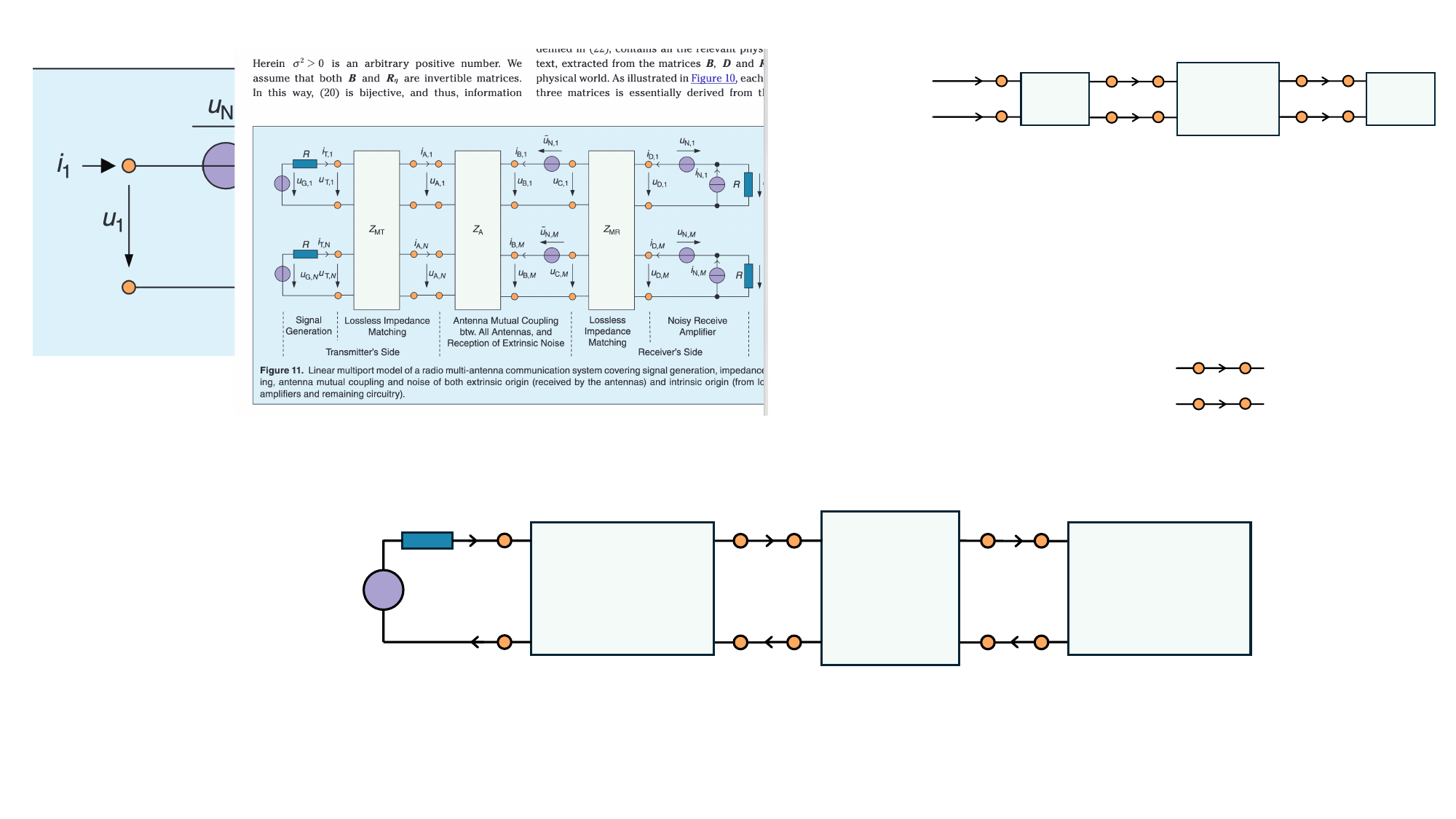}
  \caption{Multiport network model for a single pinching antenna.}
  \label{fig:single_multiport}
\end{figure*} 

Against the above background, this article aims to establish a general and physically consistent model for PASS and to explore the reconfigurability of PAs. To achieve this, we leverage the multiport network theory, which offers an equivalent circuit-theoretic framework that both ensures compliance with physical laws and preserves the applicability of signal processing and information-theoretic tools to practical communication systems \cite{6880934,10373407}. Multiport network theory has been extensively applied in wireless communications to address critical challenges such as mutual coupling \cite{ivrlavc2010toward,11006094}, reconfigurable intelligent surfaces \cite{shen2021modeling,11098513}, and super-directivity in antenna arrays \cite{9048753}. Based on this theory, the key contributions of this paper are summarized as follows:
\begin{itemize}
    \item We propose a general and physically consistent model for PASS based on the multiport network theory, which accommodates both downlink and uplink communications while accounting for potential hardware impairments. Within this model, we establish the ideal reconfigurability constraint that enables full amplitude and phase control of PAs and derive the corresponding end-to-end signal models for downlink scenarios with both single and multiple PAs.
    \item We further propose a practical reconfigurable model for PASS based on directional couplers (DCs) and derive its corresponding end-to-end signal model. This enables two practical reconfiguration mechanisms for PAs: amplitude-only control and amplitude-constrained phase control.
    \item We investigate the pinching beamforming optimization problem for both ideal and DC-based reconfigurable PAs, where PA positions and reconfigurable coefficients are jointly optimized to maximize channel gain. We derive a globally optimal solution for the ideal case and propose an efficient iterative algorithm for the practical DC-based model.
\end{itemize}

\emph{Notations:} Scalars are denoted in standard typeface, while vectors and matrices are written in lowercase boldface and uppercase boldface, respectively. The sets of complex and real numbers are denoted by $\mathbb{C}$ and $\mathbb{R}$, respectively. The inverse, conjugate, transpose, and conjugate transpose operations are represented by $(\cdot)^{-1}$, $(\cdot)^*$, $(\cdot)^T$, and $(\cdot)^H$, respectively. For a complex number $\alpha$, its amplitude and phase are denoted by $|\alpha|$ and $\angle \alpha$. The Euclidean norm is written as $\|\cdot\|$, and the real part of a complex number is denoted by $\Re\{\cdot\}$. The identity matrix is denoted by $\mathbf{I}$. A block diagonal matrix with diagonal blocks ${\mathbf{x}_1,\dots,\mathbf{x}_N}$ is represented as $\mathrm{blkdiag}(\mathbf{x}_1,\dots,\mathbf{x}_N)$. The notation $\mathcal{CN}(\mu,\sigma^2)$ denotes a circularly symmetric complex Gaussian distribution with mean $\mu$ and variance $\sigma^2$. Finally, $\mathcal{O}(\cdot)$ denotes big-O notation.  
%


\section{Results}

\subsection{Multiport Network for Single Pinching Antenna}

Fig. \ref{fig:single_multiport} illustrates the multiport network model for a system consisting of a single waveguide and a single pinching antenna (PA). The signal is initially injected into the waveguide and propagates to the right. Upon reaching the PA, part of the signal energy is coupled out of the waveguide and radiated into free space. To model this behavior, the PA is represented as a three-port network, with ports 1 and 2 corresponding to the two ends of the waveguide and port 3 opening to the wireless channel. The wireless channel between port 3 of the PA and the receiver is further modeled as a two-port network. The notations for the incident and reflected waves at each multiport interface are also defined in Fig.~\ref{fig:single_multiport}.

\subsubsection{Network Analysis}


In circuit theory, a waveguide, which is essentially a type of transmission line, can be modeled as an equivalent lumped-element network formed by an infinite concatenation of infinitesimal transmission-line sections, each described by the per-unit-length inductance $L$ and capacitance $C$. The characteristic impedance $Z_0$ and the propagation factor $\gamma_g$  of the waveguide are thus given by \cite{mongia1999rf, pozar2021microwave}
\begin{align}
    Z_0 = \sqrt{\frac{L}{C}}, \quad \gamma_g = j \omega \sqrt{LC},
\end{align}
where $\omega = 2 \pi f$ is the angular frequency and $f$ is the signal frequency. To explicitly separate the effects of attenuation and phase shift, the propagation constant is typically expressed as $\gamma_g = j \beta_g$, where $\beta_g \in \mathbb{R}$ denotes propagation constant. The propagation constant $\beta_g$ is related to the signal frequency by $\beta_g = 2\pi n_g / \lambda$, with $n_g$ denoting the effective refractive index and $\lambda$ the signal wavelength. For a lossless waveguide of length $x_g$, its behavior can be described by the following scattering matrix:
\begin{equation} \label{S_matrix_WG}
    \mathbf{T}(x_g) = \begin{bmatrix}
        0 & e^{-j\beta_g x_g} \\
        e^{-j\beta_g x_g} & 0
    \end{bmatrix},
\end{equation}
This matrix indicates that a signal entering one port exits the other port, attenuated and phase-shifted by a factor of $e^{-j\beta_g x_g}$.

The PA, modeled as a three-port network, is characterized by its scattering matrix $\mathbf{\Theta} \in \mathbb{C}^{3 \times 3}$. Let $\mathbf{a} = [a_1, a_2, a_3]^T$ and $\mathbf{b} = [b_1, b_2, b_3]^T$ denote the incident and reflected signal vectors at the PA, which are related by 
\begin{equation} \label{S_matrixA}
    \mathbf{b} = \mathbf{\Theta} \mathbf{a}.
\end{equation}  
According to \cite{pozar2021microwave}, the scattering matrix $\mathbf{\Theta}$ can be converted from impedance matrix using
\begin{equation}
\mathbf{\Theta} = \left( \mathbf{Z} + Z_0 \mathbf{I} \right)^{-1} \left( \mathbf{Z} - Z_0 \mathbf{I} \right),
\end{equation}
where $\mathbf{Z} \in \mathbb{C}^{3 \times 3}$ is the impedance matrix of the PA, and $Z_0$ is the reference impedance that is set to the characteristic impedance of the waveguide. For a passive PA, the scattering matrix must satisfy the energy conservation condition:
\begin{equation} \label{general_constraint}
\mathbf{\Theta}^H \mathbf{\Theta} \preceq \mathbf{I}.
\end{equation}

For the wireless channel between the PA and the receiver, based on the uplink/downlink reciprocity, it is represented by the following symmetric scattering matrix 
\begin{equation} \label{eq:channel_S_matrix}
    \mathbf{H} = \begin{bmatrix}
    h_{TT} & h_{TR} \\
    h_{TR} & h_{RR}
    \end{bmatrix}.
\end{equation}
In this matrix, $h_{TT}$ and $h_{RR}$ represent the reflection coefficients resulting from impedance mismatches at the radiation port (i.e., port 3) of the PA and the receiver port, while $h_{TR}$ denotes the reciprocal transmission coefficients of the wireless channel. Then, we have the following relationship:
\begin{equation} \label{eq:channel_S_relationship}
    \begin{bmatrix}
        a_3 \\ b_R
    \end{bmatrix} = \begin{bmatrix}
    h_{TT} & h_{TR} \\
    h_{TR} & h_{RR}
    \end{bmatrix} \begin{bmatrix}
        b_3 \\ a_R
    \end{bmatrix}.
\end{equation} 
The transmission coefficient $h_{TR}$ captures both the attenuation and the phase shift introduced by free-space propagation, and can be expressed as \cite{tse2005fundamentals}
\begin{equation} \label{eq:channel_model_single}
h_{TR} = \frac{\lambda}{4 \pi d} e^{-j \frac{2\pi}{\lambda} d},
\end{equation}
where $d$ denotes the distance between the PA and the receiver. In this work, multipath fading is neglected for two main reasons. First, the flexibility of PA deployment enables positioning close to the receiver, thereby ensuring a dominant LoS link. Second, in high-frequency bands, the path loss of non-line-of-sight (NLoS) components is significantly higher than that of the LoS component, rendering their contribution negligible.


\begin{remark}
    \normalfont
    Note that the scattering coefficients of the wireless channel are closely tied to the positions of the PAs, as adjusting their locations inherently changes the link distance $d$. A more detailed discussion of this relationship is provided in Section~\ref{sec:optimization}.
\end{remark}

Finally, let $Z_T$, $Z_R$, and $Z_L$ denote the load impedance at the transmitter, receiver, and the termination of the waveguide, respectively, and $a_s$ denote the source voltage at the transmitter. The corresponding incident and reflected waves at these three interfaces are related by
\begin{align}
    \label{Tx_reflection}
    a_T & = a_s + \Gamma_T b_T, \\
    \label{Rx_reflection}
    a_R & = \Gamma_R b_R,  \\
    \label{load_reflection}
    a_L & = \Gamma_L b_L,
\end{align}   
where $\Gamma_T=\frac{Z_T - Z_0}{Z_T + Z_0}$, $\Gamma_R=\frac{Z_R - Z_0}{Z_R + Z_0}$, and $\Gamma_L=\frac{Z_L - Z_0}{Z_L + Z_0}$ are the reflection coefficients of transmission, reception, and termination loads, respectively.

\subsubsection{End-to-End Signal Model}

To establish the end-to-end signal model, we need to derive the relationship between the transmit voltage $v_T$  and the receive voltage $v_R$. As illustrated in Fig.~\ref{fig:single_multiport}, these voltages are expressed in terms of the incident and reflected signal components as
\begin{align}
    v_T & = a_T + b_T, \\
    v_R & = a_R + b_R.
\end{align}

\begin{figure*}[t!]
  \centering
  \includegraphics[width=0.9\textwidth]{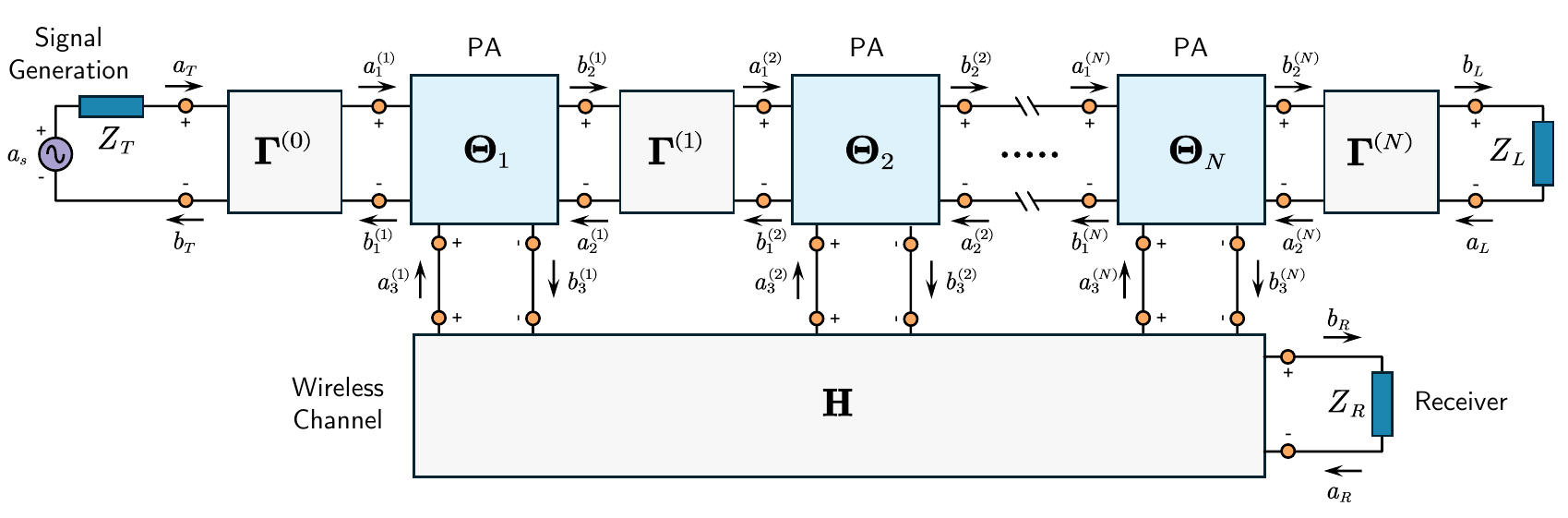}
  \caption{Multiport network model for multiple pinching antennas.}
  \label{fig:multiple_multiport}
\end{figure*} 

We begin by analyzing the three-port network of the PA. Let $x_0$ denote the length of the waveguide segment connected to port 1 of the PA. Based on the waveguide scattering matrix in \eqref{S_matrix_WG}, the incident and reflected waves at the transmitter side are related to those at port 1 of the PA by $a_T = e^{j\beta_g x_0} a_1$ and $b_T = e^{-j\beta_g x_0} b_1$. Substituting these into the transmitter reflection condition in \eqref{Tx_reflection}, we obtain 
\begin{equation} \label{port_1}
    a_1 = e^{-j\beta_g x_0} a_s + e^{-2j\beta_g x_0} \Gamma_T b_1.
\end{equation}
Similarly, for port 2 of the PA, the boundary condition at the waveguide termination yields the following expression, based on \eqref{load_reflection}:
\begin{align} \label{port_2}
    a_2 & = e^{-j2\beta_g x_1} \Gamma_L b_2, 
\end{align}
where $x_1$ denotes the length of the waveguide connected to port 2 of the PA. At port 3, the relationship between incident and reflected waves is derived from the wireless channel model in \eqref{eq:channel_S_relationship} and the receiver reflection condition in \eqref{Rx_reflection} as follows:
\begin{align}
    \label{port_3}
    a_3 & =  \left( h_{TT} + \frac{h_{TR}^2 \Gamma_R}{1 - h_{RR} \Gamma_R} \right) b_3, \\
    b_R & = \frac{h_{TR} }{1 - h_{RR} \Gamma_R} b_3.
\end{align}
Equations \eqref{port_1}-\eqref{port_3} can be compactly expressed in matrix form as
\begin{equation} \label{reflectionA}
    \mathbf{a} = e^{-j\beta_g x_0} \mathbf{e}_1 a_s + \mathbf{\Gamma} \mathbf{b},
\end{equation}
where $\mathbf{e}_i \in \mathbb{R}^{3 \times 1}$ is a vector whose $i$-th entry is one while the others are zero, and
\begin{align}
    \mathbf{\Gamma} & = \mathrm{diag} \left(e^{-j2\beta_g x_0} \Gamma_T , e^{-j2\beta_g x_1} \Gamma_L,  h_{TT} + \frac{h_{TR}^2 \Gamma_R}{1 - h_{RR} \Gamma_R} \right).
\end{align}
Combining \eqref{S_matrixA} and \eqref{reflectionA}, we can express both $\mathbf{a}$ and $\mathbf{b}$ as a function of the source voltage $a_s$ as 
\begin{align}
  \mathbf{a} & = e^{-j\beta_g x_0} (\mathbf{I} - \mathbf{\Gamma} \mathbf{\Theta})^{-1} \mathbf{e}_1 a_s, \\
  \mathbf{b} & = e^{-j\beta_g x_0} \mathbf{\Theta} (\mathbf{I} - \mathbf{\Gamma} \mathbf{\Theta})^{-1} \mathbf{e}_1 a_s,
\end{align}
where $\mathbf{e}_i \in \mathbb{R}^{3 \times 1}$ is a three-dimensional vector whose $i$-th entry is one while the others are zero.  

Based on the above analysis, the transmit voltage $v_T$ can be expressed as 
\begin{align}
    v_T = & e^{j\beta_g x_0} a_1 + e^{-j\beta_g x_0} b_1 \nonumber \\
    = & \mathbf{e}_1^T \left(\mathbf{I} + e^{-2j\beta_gx_0} \mathbf{\Theta}\right) \left( \mathbf{I} - \mathbf{\Gamma} \mathbf{\Theta} \right)^{-1} \mathbf{e}_1 a_s.
\end{align} 
The receive voltage is given by 
\begin{align}
    v_R = & (1 + \Gamma_R) b_R = \frac{(1 + \Gamma_R) h_{TR}}{1 - h_{RR} \Gamma_R} b_3\\
    = &  \frac{e^{-j\beta_g x_0} (1 + \Gamma_R) h_{TR} \mathbf{e}_3^T \mathbf{\Theta} \left( \mathbf{I} - \mathbf{\Gamma} \mathbf{\Theta} \right)^{-1} \mathbf{e}_1 }{1 - h_{RR} \Gamma_R} a_s
\end{align}
The ratio of receive to transmit voltages is thus given by 
\begin{align} \label{eq:general_single_model}
    \frac{v_R}{v_T} = \frac{e^{-j\beta_g x_0} (1 + \Gamma_R) h_{TR} \mathbf{e}_3^T \mathbf{\Theta} \left( \mathbf{I} - \mathbf{\Gamma} \mathbf{\Theta} \right)^{-1} \mathbf{e}_1 }{ \left(1 - h_{RR} \Gamma_R\right) \mathbf{e}_1^T \left(\mathbf{I} + e^{-2j\beta_gx_0} \mathbf{\Theta}\right) \left( \mathbf{I} - \mathbf{\Gamma} \mathbf{\Theta} \right)^{-1} \mathbf{e}_1}.
\end{align}

The result in \eqref{eq:general_single_model} provides a general end-to-end relationship (excluding the additive white Gaussian noise at the receiver), capturing the effects of impedance mismatches at the transmitter, receiver, and termination load, as well as the wireless channel. However, this general model is analytically complex and offers limited insight into the role of the PA in communication performance. Therefore, in the following analysis, we consider a special case with perfect impedance matching, i.e., $\Gamma_T = \Gamma_R = \Gamma_L = h_{TT} = h_{RR} = 0$. Under these assumptions, \eqref{eq:general_single_model} can be simplified into 
\begin{equation}
    \frac{v_R}{v_T} = \frac{e^{-j\beta_g x_0} h_{TR} \Theta_{31}}{1 + e^{-j2\beta_g x_0}\Theta_{11}},
\end{equation}
where $\Theta_{ij}$ denote the entry in the $i$-th row and $j$-th column of $\mathbf{\Theta}$.   
Therefore, the simplified end-to-end noisy signal model can be expressed as 
\begin{equation}
    y = \frac{e^{-j\beta_g x_0} h_{TR} \Theta_{31}}{1 + e^{-j2\beta_g x_0}\Theta_{11}} s + w,
\end{equation}
where $w \sim \mathcal{CN}(0,\sigma_w^2)$ denotes the additive white Gaussian noise, and $s \in \mathbb{C}$ and $y \in \mathbb{C}$ are the transmit and receive signals, respectively.

\subsection{Multiport Network for Multiple Pinching Antennas}

In this section, we extend the multiport network model to the case of multiple PAs attached to the waveguide, as illustrated in Fig. \ref{fig:multiple_multiport}. The scattering matrices of the $n$-th PA is denoted by $\mathbf{\Theta}_n \in \mathbb{C}^{3 \times 3}$. The notations for the incident and reflected waves at each multiport interface follow those defined in Fig. \ref{fig:single_multiport}. For clarity and conciseness, we merge the notations at the interfaces to simplify the overall representation.

\subsubsection{Network Analysis} 
Based on the definition of the scattering matrix, we have the following relationship for each PA:
\begin{equation}
    \mathbf{b}_n = \mathbf{\Theta}_n \mathbf{a}_n,  n=1,\dots,N,
\end{equation}
where $\mathbf{a}_n = [a_{1}^{(n)}, a_{2}^{(n)}, a_{3}^{(n)}]^T$ and $\mathbf{b}_n = [b_{1}^{(n)}, b_{2}^{(n)}, b_{3}^{(n)}]^T$. The above relationships can be collectively written as 
\begin{equation}
    \label{eq:S_matirx_multiple}
    \mathbf{b} = \mathbf{S} \mathbf{a},
\end{equation}
where $\mathbf{a} = [\mathbf{a}_1^{T},\dots,\mathbf{a}_N^{T}]^T$, $\mathbf{b} = [\mathbf{b}_1^{T},\dots,\mathbf{b}_N^{T}]^T$, and $\mathbf{S} = \mathrm{blkdiag}(\mathbf{\Theta}_1,\dots,\mathbf{\Theta}_N)$. 

The main challenge in analyzing the multiport network with multiple PAs lies in handling the cascaded configuration of the three-port networks. To characterize their collective behavior, we model the cascade of $N$ three-port PAs as a single $(N+2)$-port network. In this formulation, the ports of each PA are categorized as either \emph{external} or \emph{internal}:
\begin{itemize}
    \item For the first PA ($n=1$), ports 1 and 3 are external, while port 2 serves as an internal connection to the second PA. Similarly, for the last PA ($n=N$), ports 2 and 3 are external, with port 1 acting as the internal connection. 
    \item For each middle PA ($n=2,\dots,N-1$), only port 3 is external, whereas ports 1 and 2 are internal, connected to adjacent PAs.
\end{itemize}

Based on the above classification, we rewrite \eqref{eq:S_matirx_multiple} as
\begin{equation}
    \label{eq:cascaded_open}
    \begin{bmatrix}
        \mathbf{b}_E \\ \mathbf{b}_I
    \end{bmatrix} = 
    \begin{bmatrix}
        \mathbf{S}_{EE} & \mathbf{S}_{EI} \\ \mathbf{S}_{IE} & \mathbf{S}_{II}
    \end{bmatrix} \begin{bmatrix}
        \mathbf{a}_E \\ \mathbf{a}_I
    \end{bmatrix},
\end{equation}
where  $\mathbf{a}_E \in \mathbb{C}^{(N+2)\times 1}$ and $\mathbf{b}_E \in \mathbb{C}^{(N+2)\times 1}$ denote the incident and reflected waves at \emph{external} ports, respectively, and $\mathbf{a}_I \in \mathbb{C}^{2(N-1)\times 1}$ and $\mathbf{b}_I \in \mathbb{C}^{2(N-1)\times 1}$ denote the incident and reflected waves at \emph{internal} ports, respectively. Specifically, the vectors $\mathbf{a}_E$ and $\mathbf{a}_I$   are defined as 
\begin{align}
    \mathbf{a}_E & = \Big[ a_{1}^{(1)}, a_{3}^{(1)}, \underbrace{a_{3}^{(2)},\dots,a_{3}^{(N-1)},}_{\text{Port 3 of the middle PAs}} a_{3}^{(N)}, a_{2}^{(N)}  \Big]^T, \\
    \mathbf{a}_I & = \Big[ a_{2}^{(1)}, \underbrace{a_{1}^{(2)}, a_{2}^{(2)},\dots,a_{1}^{(N-1)}, a_{2}^{(N-1)}}_{\text{Ports 1 and 2 of the middle PAs}} a_{1}^{(N)}  \Big]^T, 
\end{align}
and $\mathbf{b}_E$ and $\mathbf{b}_I$ are defined analogously. Based on the above arrangement, the submatrices $\mathbf{S}_{EE}$, $\mathbf{S}_{EI}$, $\mathbf{S}_{IE}$, and $\mathbf{S}_{II}$ can be constructed from \eqref{eq:S_matirx_multiple} by appropriately re-indexing the rows and columns of the overall scattering matrix $\mathbf{S}$. 

The expression in \eqref{eq:cascaded_open} describes the relationship between external and internal ports under the assumption that the internal ports are unconnected. Once the internal ports are connected via the TL segments, each internal port's outgoing signal becomes the incident signal to the next block, and vice versa. More particularly, given the scattering matrix $\mathbf{T}(x_n)$ of the waveguide between the $n$-th PA and the $(n+1)$-th PA, the internal connection between adjacent PAs is described by 
\begin{equation}
    \begin{bmatrix}
        a_{2}^{(n)} \\ a_{1}^{(n+1)}
    \end{bmatrix} = \mathbf{T}(x_n) \begin{bmatrix}
        b_{2}^{(n)} \\ b_{1}^{(n+1)}
    \end{bmatrix},  n = 1,\dots,(N-1).
\end{equation}
This internal coupling relationship can be compactly expressed in matrix form as
\begin{equation}
    \label{eq:cascaded_connected}
    \mathbf{a}_I = \mathbf{T}_I \mathbf{b}_I,
\end{equation}
where $\mathbf{T}_I$ is a block-diagonal matrix defined as 
\begin{equation}
    \mathbf{T}_I = \mathrm{blkdiag} \left( \mathbf{T}(x_1),\dots,\mathbf{T}(x_{N-1})  \right).
\end{equation}
Combining \eqref{eq:cascaded_open} and \eqref{eq:cascaded_connected}, we obtain the following scattering relationship of the external ports:
\begin{align} \label{external_scattering}
    \mathbf{b}_E & = \mathbf{\Phi} \mathbf{a}_E, \\
    \mathbf{\Phi} & = \mathbf{S}_{EE} + \mathbf{S}_{EI}(\mathbf{I} - \mathbf{T}_I \mathbf{S}_{II})^{-1} \mathbf{T}_I \mathbf{S}_{IE}.
\end{align}

For the wireless channel involving multiple PAs, the corresponding scattering matrix is given by
\begin{equation}
    \mathbf{H} = \begin{bmatrix}
        \mathbf{H}_{TT} & \mathbf{h}_{TR} \\
        \mathbf{h}_{TR}^T & h_{RR}
    \end{bmatrix},
\end{equation} 
where $\mathbf{H}_{TT} \in \mathbb{C}^{N \times N}$ captures the reflection and mutual coupling coefficients among the $N$ PAs, $h_{RR}$ denotes the reflection coefficients at the receiver, and $\mathbf{h}_{TR} \in \mathbb{C}^{N \times 1}$ contains the transmission coefficients from each PA to the receiver. Let $\mathbf{a}_3 = [a_3^{(1)},\dots,a_3^{(N)}]^T$ and $\mathbf{b}_3 = [b_3^{(1)},\dots,b_3^{(N)}]^T$ denote the incident and reflected signal vectors at the radiation ports of the $N$ PAs, respectively. The input-output relationship for the wireless channel is then given by
\begin{equation} \label{eq:channel_S_relationship_multiple}
    \begin{bmatrix}
        \mathbf{a}_3 \\ b_R
    \end{bmatrix} = \begin{bmatrix}
        \mathbf{H}_{TT} & \mathbf{h}_{TR} \\
        \mathbf{h}_{TR}^T & h_{RR}
    \end{bmatrix} \begin{bmatrix}
        \mathbf{b}_3 \\ a_R
    \end{bmatrix}.
\end{equation}
Similar to the single-PA case in \eqref{eq:channel_model_single}, the $n$-th entry of $\mathbf{h}_{TR}$, denoted by $h_{TR,n}$, is modeled as
\begin{align} \label{wireless_channel_coefficients}
    h_{TR,n} = \frac{\lambda}{4 \pi d_n} e^{-j \frac{2\pi}{\lambda} d_n},
\end{align}   
where $d_n$ is the distance from the $n$-th PA to the receiver.

Finally, the boundary conditions at the transmitter, receiver, and termination load remain the same as \eqref{Tx_reflection}-\eqref{load_reflection}.

\subsubsection{End-to-End Signal Model} We now derive the relationship between the transmit voltage $v_T = a_T + b_T$ and the receive voltage $v_R = a_R + b_R$  to establish the end-to-end signal model for the multi-PA system. Following the single-PA case in \eqref{port_1} and \eqref{port_2}, the boundary conditions at the transmitter and termination load are given by
\begin{align}
    \label{port_1_multiple}
    a_1^{(1)} & = e^{-j\beta_g x_0} a_s + e^{-2j\beta_g x_0} \Gamma_T b_1^{(1)}, \\
    \label{port_2_multiple}
    a_2^{(N)} & = e^{-2j\beta_g x_N} \Gamma_L b_2^{(N)}.
\end{align}
At port 3 of all PAs, using the multi-PA wireless channel model in \eqref{eq:channel_S_relationship_multiple} and the receiver boundary condition in \eqref{Rx_reflection}, we obtain
\begin{align}
    \label{port_3_multiple}
    \mathbf{a}_3 & = \left( \mathbf{H}_{TT} + \frac{\mathbf{h}_{TR} \mathbf{h}_{TR}^T \Gamma_R}{1 - h_{RR} \Gamma_R} \right) \mathbf{b}_3, \\
    \label{port_3_multiple_2}
    b_R & = \frac{\mathbf{h}_{TR}^T}{1 - h_{RR} \Gamma_R} \mathbf{b}_3.
\end{align}
Equations \eqref{port_1_multiple}-\eqref{port_3_multiple} can be written in matrix form as
\begin{equation} \label{reflectionB}
    \mathbf{a}_E = e^{-j\beta_g x_0} \mathbf{u}_1 a_s + \mathbf{\Sigma} \mathbf{b}_E,
\end{equation}
where $\mathbf{u}_i \in \mathbb{R}^{(N+2) \times 1}$ is a $(N+2)$-dimensional unit vector with a one in the $i$-th position and zeros elsewhere, and
\begin{equation}
    \mathbf{\Sigma} = \begin{bmatrix}
        e^{-2j\beta_g x_0} \Gamma_T & \mathbf{0} & 0 \\
        0 & \mathbf{H}_{TT} + \frac{\mathbf{h}_{TR} \mathbf{h}_{TR}^T \Gamma_R}{1 - h_{RR} \Gamma_R} & 0 \\
        0 & \mathbf{0} & e^{-2j\beta_g x_N} \Gamma_L
    \end{bmatrix}.
\end{equation}
Substituting the external scattering relation in \eqref{external_scattering} into \eqref{reflectionB}, we obtain
\begin{align}
  \mathbf{a}_E & = e^{-j\beta_g x_0} (\mathbf{I} - \mathbf{\Sigma} \mathbf{\Phi})^{-1} \mathbf{u}_1 a_s, \\
  \mathbf{b}_E & = e^{-j\beta_g x_0} \mathbf{\Phi} (\mathbf{I} - \mathbf{\Sigma} \mathbf{\Phi})^{-1} \mathbf{u}_1 a_s.
\end{align}

Based on the above analysis, the transmit voltage $v_T$ can be expressed as 
\begin{align}
    v_T = & e^{j\beta_g x_0} a_1^{(1)} + e^{-j\beta_g x_0} b_1^{(1)} \nonumber \\
    = & \mathbf{u}_1^T \left(\mathbf{I} + e^{-2j\beta_gx_0} \mathbf{\Phi}\right) \left( \mathbf{I} - \mathbf{\Sigma} \mathbf{\Phi} \right)^{-1} \mathbf{u}_1 a_s.
\end{align} 

Furthermore, based on the relationship in \eqref{port_3_multiple_2} and the fact that $\mathbf{b}_3$ is a subvector of $\mathbf{b}_E$ spanning entries 2 to $N$, the receive voltage can be calculated as 
\begin{align}
    v_R = & (1 + \Gamma_R) b_R = \frac{(1 + \Gamma_R) \mathbf{h}_{TR}^T}{1 - h_{RR} \Gamma_R} \mathbf{b}_3\\
    = &  \frac{e^{-j\beta_g x_0} (1 + \Gamma_R) \mathbf{g}_{TR}^T \mathbf{\Phi} (\mathbf{I} - \mathbf{\Sigma} \mathbf{\Phi})^{-1} \mathbf{u}_1 }{1 - h_{RR} \Gamma_R} a_s,
\end{align}
where $\mathbf{g}_{TR} = [0, \mathbf{h}_{TR}^T, 0]^T$. Consequently, the transmit-receive voltage ratio is given by
\begin{equation}
    \frac{v_R}{v_T} = \frac{e^{-j\beta_g x_0} (1 + \Gamma_R) \mathbf{g}_{TR}^T \mathbf{\Phi} (\mathbf{I} - \mathbf{\Sigma} \mathbf{\Phi})^{-1} \mathbf{u}_1 }{\left(1 - h_{RR} \Gamma_R\right)\mathbf{u}_1^T \left(\mathbf{I} + e^{-2j\beta_gx_0} \mathbf{\Phi}\right) \left( \mathbf{I} - \mathbf{\Sigma} \mathbf{\Phi} \right)^{-1} \mathbf{u}_1 }
\end{equation}

For the special case where there is no mutual coupling and perfect impedance matching, i.e., $\Gamma_T = \Gamma_R = \Gamma_L = h_{RR} = 0$ and $\mathbf{H}_{TT} = \mathbf{0}$, the expression simplifies to
\begin{align}
    \frac{v_R}{v_T} = & \frac{e^{-j\beta_g x_0}  \mathbf{g}_{TR}^T \mathbf{\Phi} \mathbf{u}_1 }{ 1 + e^{-2j\beta_gx_0} \mathbf{u}_1^T \mathbf{\Phi} \mathbf{u}_1 } =  \frac{e^{-j\beta_g x_0}  \mathbf{h}_{TR}^T \boldsymbol{\phi}_T}{ 1 + e^{-2j\beta_gx_0} \phi_R },
\end{align}
where $\boldsymbol{\phi}_T = \mathbf{\Phi}(2:N+1, 1)$ and $\phi_R = \mathbf{\Phi}(1,1)$ represent the effective transmission and reflection coefficients of the cascaded PA system under these assumptions.
Accordingly, the simplified end-to-end noisy signal model is given by
\begin{equation} \label{multi_antenna_signal}
    y = \frac{e^{-j\beta_g x_0}  \mathbf{h}_{TR}^T \boldsymbol{\phi}_T}{ 1 + e^{-2j\beta_gx_0} \phi_R } s + w.
\end{equation}

\begin{remark}
    \normalfont
    \emph{(Generality of the Proposed Model)} The generality of the proposed model for PASS is reflected in two aspects. On the one hand, it can be directly extended to uplink scenarios by placing the voltage source at the receiver and following the same analytical procedure. On the other hand, impedance mismatches caused by hardware impairments are naturally incorporated through the reflection coefficients within the model.
\end{remark}

\begin{remark}
    \normalfont
    \emph{(Ideal Reconfigurability of PAs)}
    In the ideal case, the reconfigurability of the PAs can be realized by tuning their scattering matrices $\mathbf{\Theta}_n$, subject to the general energy conservation constraint  $\mathbf{\Theta}_n^H \mathbf{\Theta}_n \preceq \mathbf{I}$ as specified in \eqref{general_constraint}. These ideal reconfigurable PAs can be modeled as impedance networks $\mathbf{Z}_n \in \mathbb{C}^{3 \times 3}$ with arbitrarily adjustable impedance values, resulting in a fully reconfigurable scattering matrix given by $\mathbf{\Theta}_n = (\mathbf{Z}_n + Z_0 \mathbf{I})^{-1}(\mathbf{Z}_n - Z_0 \mathbf{I})$. It is important to note that the energy conservation constraint only requires the total signal power at all ports not to exceed the input power. Within this feasible region, both signal amplitude and phase can be freely adjusted, thereby granting each PA full reconfiguration capability.
\end{remark}

\begin{remark}
    \normalfont
    \emph{(Enhanced Pinching Beamforming)} In conventional PASS, pinching beamforming refers to adjusting the positions of PAs along the waveguide so that their signals combine constructively at the intended communication users \cite{ding2025flexible,wang2025modeling}. This relies on the fact that repositioning the PAs alters the scattering matrices of the corresponding waveguide segments and also the wireless channels, thereby modifying the signal phases related each PA. In our proposed reconfigurable PASS, the pinching beamforming capability is further enhanced by allowing the scattering matrices of individual PAs to be reconfigured, thereby providing additional amplitude and phase control capability.
\end{remark}

\subsection{Directional Coupler for Pinching Antennas}
In the previous section, we introduced the loose energy conservation constraints associated with ideal PA reconfigurability. However, in practical scenarios, the reconfigurability of a PA is constrained by the limitations of its physical implementation. In the following, we introduce a practical PA design based on a reconfigurable DC. This implementation is motivated by three key advantages. First, DCs can be realized via near-field evanescent coupling between two adjacent waveguides, i.e., transmission lines. Leveraging this principle, a PA can be implemented by placing a short waveguide in proximity to the main waveguide, offering a low-cost hardware solution. Second, the coupling is contactless and requires no permanent or invasive electrical connection, enabling flexible attachment, detachment, and position adjustment of the PA along the main waveguide. Finally, the reconfigurability of the DC can be readily achieved by adjusting the coupling gap between the PA and the main waveguide using microelectromechanical systems (MEMS)~\cite{6568978}.

\subsubsection{Scattering Matrix Characterization}
\begin{figure}[t!]
  \centering
  \includegraphics[width=0.48\textwidth]{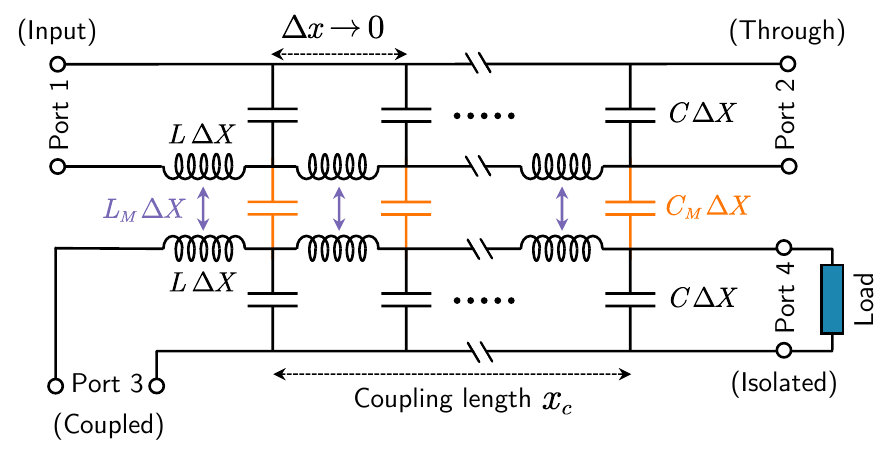}
  \caption{Lumped equivalent circuit of the DC-based pinching antenna.}
  \label{fig:directional_coupler}
\end{figure}

Fig. \ref{fig:directional_coupler} illustrates the lumped equivalent circuit of the DC-based PA. 
In particular, the behavior of the coupler can be characterized by three sets of parameters: 1) the intrinsic parameters of each waveguide, given by the per-unit inductance $L$ and capacitance $C$, 2) the coupling parameters between the waveguides, represented by the per-unit mutual inductance $L_M$ and capacitance $C_M$, and 3) the coupling length $x_c$. To analyze DCs, the odd/even mode analysis is typically applied. In particular, the DC can be represented by a linear superposition of an even mode coupler and an odd mode coupler, whose characteristic impedance are respectively given by \cite{mongia1999rf, pozar2021microwave} 
\begin{align} \label{even_odd_impedance}
    Z_0^e = \sqrt{\frac{L+L_M}{C-C_M}}, \quad Z_0^o = \sqrt{\frac{L-L_M}{C+C_M}}.
\end{align}
Their propagation constants are respectively given by 
\begin{align}
    &\beta_e = \omega \sqrt{(L + L_M)(C - C_M)}, \\
    &\beta_o = \omega \sqrt{(L - L_M)(C + C_M)}.
\end{align}

Note that the DC is inherently a four-port network. In the following, we first discuss its four-port scattering matrix, denoted by $\mathbf{\Xi}\in \mathbb{C}^{4 \times 4}$, and then explain how it reduces to a three-port network in the context of PAs. Let $\Xi_{nm}$ denote the scattering parameters in $\mathbf{\Xi}$ from the $n$-th port to the $m$-th port. An important design criterion of a DC is that each port is matched, i.e., $\Xi_{11}=\Xi_{22}=\Xi_{33}=\Xi_{44} = 0$, which is satisfied if the even and odd mode characteristic impedances are related as \cite{mongia1999rf, pozar2021microwave}
\begin{equation} \label{matching_comdition}
    Z_0^e Z_0^o = Z_0^2. 
\end{equation}
Substituting \eqref{even_odd_impedance} into \eqref{matching_comdition} yields
\begin{align}
    &\sqrt{\frac{L^2-L_M^2}{C^2-C_M^2}} = Z_0^2 \, \Rightarrow \,  \sqrt{\frac{1 - \left(\frac{L_M}{L}\right)^2}{1 - \left(\frac{C_M}{C}\right)^2}} = 1.
\end{align}
It can be readily observed that the above condition is met if 
\begin{equation}
    \frac{L_M}{L} = \frac{C_M}{C}.
\end{equation}
Subsequently, it can be proved that the propagation constants of the even and odd modes are identical, i.e.,
\begin{equation}
    \beta_e = \beta_o = \omega \sqrt{LC - L_M C_M} \triangleq \beta_c.
\end{equation}
Under the above matched conditions, the remaining scattering coefficients in $\mathbf{\Xi}$ are given by \cite{mongia1999rf, pozar2021microwave} 
\begin{align}
    &\Xi_{14} = \Xi_{41} = \Xi_{23} = \Xi_{32} = 0,\\
    \label{DC_coupling_coefficient_1}
    &\Xi_{12} = \Xi_{21} = \Xi_{34} = \Xi_{43} \nonumber \\
    & \hspace{0.55cm}= \frac{\sqrt{1 - \kappa^2}}{\sqrt{1 - \kappa^2} \cos \varphi + j \sin \varphi} \triangleq \widetilde{\Theta}_1 (\kappa), \\
    \label{DC_coupling_coefficient_2}
    &\Xi_{13} = \Xi_{31} = \Xi_{24} = \Xi_{42} \nonumber \\
    & \hspace{0.55cm}= \frac{j \kappa \sin \varphi}{\sqrt{1 - \kappa^2} \cos \varphi + j \sin \varphi} \triangleq \widetilde{\Theta}_2 (\kappa)
\end{align}  
where  $\varphi \triangleq \beta_c x_c$ and $\kappa$ is the coupling coefficient, defined by 
\begin{equation} \label{DC_coupling_coefficients}
    \kappa = \frac{Z_0^e - Z_0^o}{Z_0^e + Z_0^o}.
\end{equation}
According to \eqref{even_odd_impedance} and \eqref{DC_coupling_coefficients}, the coupling coefficient spans the range $\kappa \in [0,1)$, where $\kappa=0$ corresponds to $L_M=C_M=0$ and $\kappa \rightarrow 1$ occurs when $Z_0^e \rightarrow \infty$.

\begin{figure*}[t!]
  \centering
  \includegraphics[width=0.4\textwidth]{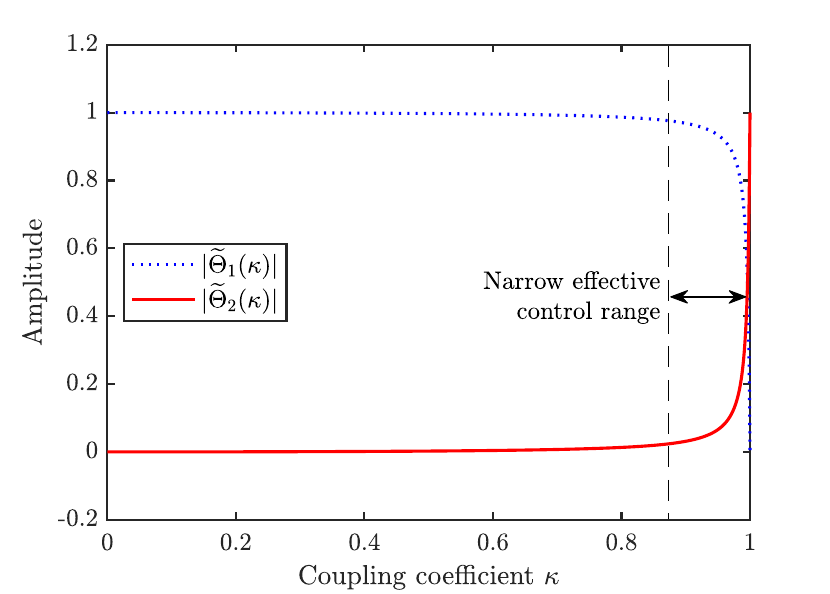}
  \includegraphics[width=0.4\textwidth]{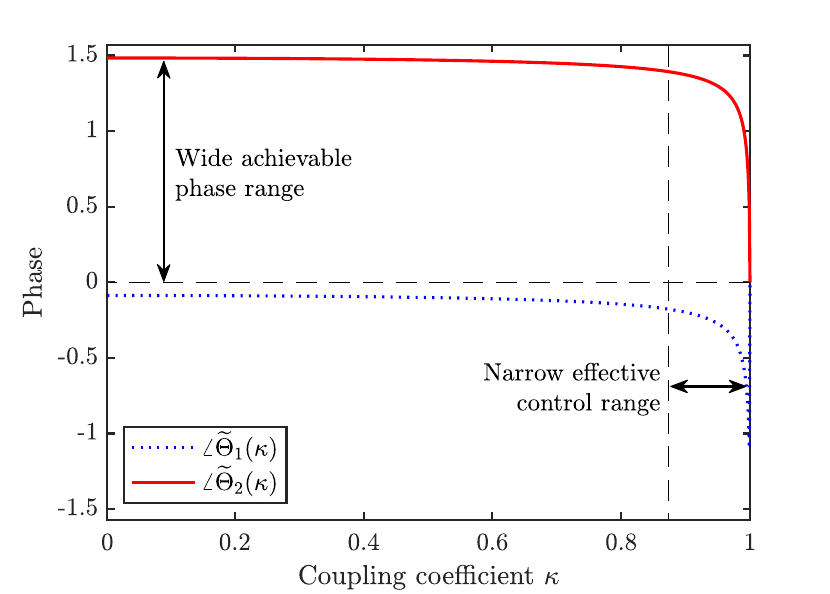}
  \caption{Amplitude-constrained phase control with $\varphi = 5^{\degree}$. A wide achievable phase range is realized at the cost of narrow effective control range of the coupling coefficient $\kappa$.} \label{Fig_ACPA_1}

  \includegraphics[width=0.4\textwidth]{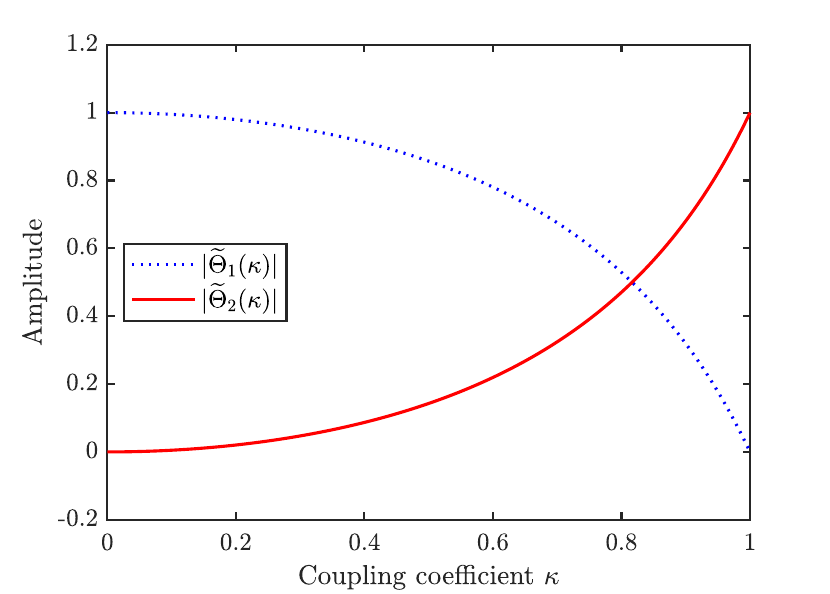}
  \includegraphics[width=0.4\textwidth]{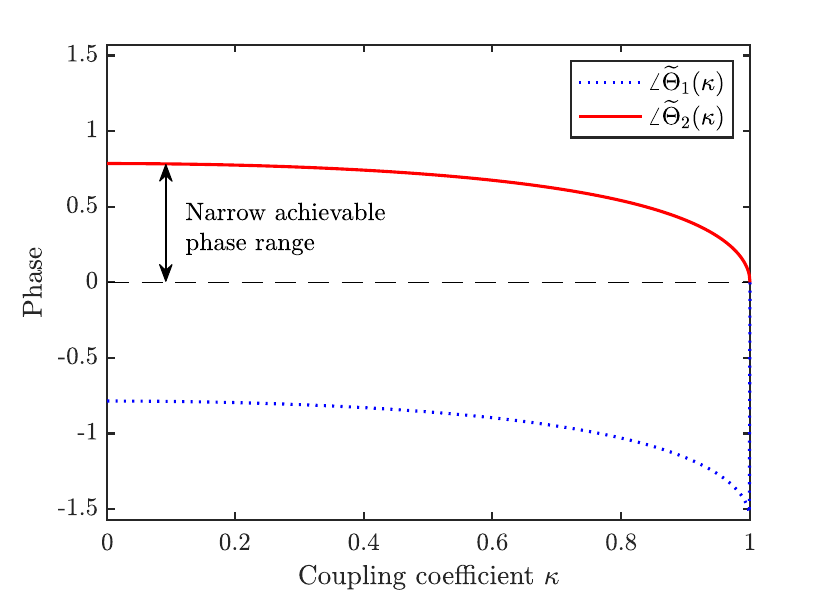}
  \caption{Amplitude-constrained phase control with $\varphi = 45^{\degree}$. Only a narrow achievable phase range is realized, but the effective control range spans almost the entire range of $\kappa \in [0,1)$.} \label{Fig_ACPA_2}
\end{figure*} 

Based on the above analysis, in the considered system, port 1 serves as the input port, while port 4 is isolated and treated as a dummy port. Consequently, only the coefficients between ports 1, 2, and 3 are relevant, yielding the following three-port scattering matrix for the $n$-th DC-based PA:
\begin{equation} \label{DC_S_matrix}
    \mathbf{\Theta}_n = \begin{bmatrix}
        0 & \widetilde{\Theta}_1(\kappa_n) & \widetilde{\Theta}_2(\kappa_n) \\
        \widetilde{\Theta}_1(\kappa_n) & 0 & 0 \\
        \widetilde{\Theta}_2(\kappa_n) & 0 & 0
    \end{bmatrix}.
\end{equation}
Here, $\kappa_n$ denotes the coupling coefficient of the $n$-th PA. 

This new scattering matrix substantially simplifies the signal model for multiple PAs. In particular, because all ports are matched, signal propagation in the waveguide occurs unidirectionally without reflections. Consequently, the overall multi-PA channel in \eqref{multi_antenna_signal} reduces to
\begin{align} \label{DC_overall_channel}
    &\frac{e^{-j\beta_g x_0}  \mathbf{h}_{TR}^T \boldsymbol{\phi}_T}{ 1 + e^{-2j\beta_gx_0} \phi_R } \nonumber \\ 
    &\hspace{0.5cm}= \sum_{n=1}^N h_{TR,n}  \left( \widetilde{\Theta}_2(\kappa_n) \prod_{i=1}^{n-1} \widetilde{\Theta}_1(\kappa_i) \right) e^{-j\beta_g \sum_{i=0}^{n-1} x_i},
\end{align} 
where the terms on the left-hand side, middle, and right-hand side correspond to the wireless channel, the cumulative PA reconfiguration coefficients, and the in-waveguide propagation, respectively.

\subsubsection{Reconfigurability Analysis}

As discussed previously, the reconfigurability of DC-based PAs can be realized by adjusting the gap between the PA and the main waveguide using MEMS, which effectively tunes the coupling coefficient $\kappa$. Following \eqref{DC_coupling_coefficient_1} and \eqref{DC_coupling_coefficient_2}, this tuning simultaneously reconfigures both the amplitude and phase of the signals radiated from the DC-based PAs. To further elaborate on this mechanism, we analyze the amplitude and phase characteristics of the corresponding scattering coefficients in the following.

\begin{table*}
    \centering
    \caption{Comparison of the Ideal and the DC-based Pinching Antennas.}
    \label{table_comparison}
    \begin{tabular}{c|c|c}
        \Xhline{1pt}
        \textbf{Model} & \textbf{Reconfigurable Coefficients} & \textbf{Reconfigurability} \\
        \hline
        Ideal PA & $\mathbf{\Theta}^H \mathbf{\Theta} \preceq \mathbf{I}$  & Full amplitude and phase control  \\ 
        \hline
        \multirow{3}{*}{DC-based PA} & \multirow{2}{*}{$\kappa \in [0,1) \Rightarrow \begin{cases}
            \widetilde{\Theta}_1 (\kappa) = \frac{\sqrt{1 - \kappa^2}}{\sqrt{1 - \kappa^2} \cos \varphi + j \sin \varphi} \\
            \widetilde{\Theta}_2 (\kappa) = \frac{j \kappa \sin \varphi}{\sqrt{1 - \kappa^2} \cos \varphi + j \sin \varphi}
            \end{cases}$}  & \makecell[c]{Amplitude-only control\\($\varphi = \pi/2$)} \\ \cline{3}
            & & \makecell[c]{Amplitude-constrained phase control \\ ($\varphi \neq \pi/2$)} \\
        \Xhline{1pt}
    \end{tabular}
\end{table*} 

The amplitudes and phases of $\widetilde{\Theta}_1(\kappa)$ and $\widetilde{\Theta}_2(\kappa)$ are
\begin{align}
    &| \widetilde{\Theta}_1(\kappa) | = \sqrt{\frac{1 - \kappa^2}{1 - \kappa^2 \cos^2 \varphi}}, \\ 
    \label{coupled_port_amplitude}
    &| \widetilde{\Theta}_2(\kappa) | = \sqrt{\frac{\kappa^2 \sin^2 \varphi}{1 - \kappa^2 \cos^2 \varphi}}, \\
    &\angle \widetilde{\Theta}_1(\kappa) = \tan^{-1} \left( \frac{-\tan \varphi}{\sqrt{1-\kappa^2}} \right), \\
    &\angle \widetilde{\Theta}_2(\kappa) = \tan^{-1} \left( \frac{\sqrt{1 - \kappa^2}}{\tan \varphi} \right).
\end{align}
It can be easily proved that the following relationships hold:
\begin{align}
    &| \widetilde{\Theta}_1(\kappa) |^2 + | \widetilde{\Theta}_2(\kappa) |^2  = 1, \\
    \label{phase_relationship}
    &\angle \widetilde{\Theta}_2(\kappa) - \angle \widetilde{\Theta}_1(\kappa)  = \frac{\pi}{2}.
\end{align}
By pre-configuring $\varphi$ to different values through the design of an appropriate coupling length $x_c$, different forms of reconfigurability can be achieved. In the following, we discuss several promising choices of $\varphi$:
\begin{itemize}
    \item \textbf{Amplitude-only control (AOC):} Choose $\varphi = \pi/2$. In this case, the scattering coefficients reduce to 
    \begin{equation}
        \widetilde{\Theta}_1(\kappa) = -j\sqrt{1-\kappa^2}, \quad \widetilde{\Theta}_2(\kappa) = \kappa,
    \end{equation}
    which implies that the phases remain constant while only the amplitudes can be tuned through variations in the coupling coefficient $\kappa$.
    \item \textbf{Amplitude-constrained phase control (ACPC):} Choose $\varphi \neq \pi/2$. In this case, signal phase control becomes possible, but it is strongly coupled with amplitude control. We first examine the phase aspect, where the key question is the extent of the achievable phase range. Based on the monotonicity of $1/\tan(\varphi)$ function for $\varphi \in (0, \pi)$, the range of achievable $\angle \widetilde{\Theta}_2(\kappa)$ by adjusting $\kappa \in [0,1)$ is given by 
    \begin{equation}
        \angle \widetilde{\Theta}_2(\kappa) = \begin{cases}
            (0, \frac{\pi}{2} - \varphi], & \text{if } \varphi \in (0, \frac{\pi}{2}), \\
            [\frac{\pi}{2} - \varphi, 0), & \text{if } \varphi \in (\frac{\pi}{2}, \pi).
        \end{cases}
    \end{equation} 
    Therefore, the tunable span of the phase is given by $|\tfrac{\pi}{2}-\varphi|$. From the relationship in \eqref{phase_relationship}, the same span also applies to $\angle \widetilde{\Theta}_1(\kappa)$. This result suggests that choosing $\varphi$ close to $0$ or $\pi$ maximizes the achievable phase range. However, such a choice is not always desirable in practice. The reason is that when $\varphi$ approaches $0$ or $\pi$, we have $\sin\varphi\to 0$ and $\cos\varphi\to 1$. In this regime, the amplitudes $|\widetilde{\Theta}_1(\kappa)|$ and $|\widetilde{\Theta}_2(\kappa)|$ become extremely insensitive for large $\kappa$, thereby restricting the effective control of $\kappa$ to a very narrow range, as depicted in Fig. \ref{Fig_ACPA_1}. Consequently, extremely precise control of $\kappa$ is required, which translates into higher hardware cost and complexity.  By contrast, when $\varphi$ approaches $\pi/2$, both amplitudes and phases vary smoothly with $\kappa$, enabling an almost full effective control range, but also leading to a narrow achievable phase range, as shown in Fig. \ref{Fig_ACPA_2}. These observations highlight a fundamental performance-complexity tradeoff for ACPC.
\end{itemize}
 
Table \ref{table_comparison} summarizes the key coefficients and distinguishing features of the ideal and DC-based PAs.

\subsection{Reconfigurable Pinching Beamforming Optimization} \label{sec:optimization}

Accordingly, we aim to jointly optimize the PA positions and their reconfigurable scattering matrices to maximize pinching beamforming performance.

To facilitate optimization, we establish a three-dimensional Cartesian coordinate system for the considered scenario. Without loss of generality, we assume the waveguide is deployed along the $x$-axis with fixed $y$- and $z$-coordinates denoted by $y_g$ and $z_g$, respectively. The coordinate of the $n$-th PA is then $ \mathbf{p}_n = ( \sum_{i=0}^n x_i, y_g, z_g )$. Let the receiver position be $\mathbf{p}_r = (x_r, y_r, z_r)$. The distance from the $n$-th PA to the receiver is
\begin{equation} \label{expression_d}
    d_n = \|\mathbf{p}_n - \mathbf{p}_r\| = \sqrt{\left(\sum_{i=0}^{n-1} x_i - x_r\right)^2 + \xi  },
\end{equation}  
where $\xi = (y_g-y_r)^2 + (z_g-z_r)^2$. The wireless channel can then be characterized as in \eqref{wireless_channel_coefficients}. From the signal model in \eqref{multi_antenna_signal}, maximizing the received power is equivalent to maximizing the overall channel gain:
\begin{equation}
    \mathcal{H}(\mathbf{x}, \widetilde{\mathbf{\Theta}}) = \left|\frac{e^{-j\beta_g x_0}  \mathbf{h}_{TR}^T \boldsymbol{\phi}_T}{ 1 + e^{-2j\beta_gx_0} \phi_R } \right|^2,
\end{equation}
where $\mathbf{x} = [x_0,\dots,x_N]$ denotes the lengths of the waveguide segments between PAs and $\widetilde{\mathbf{\Theta}} = \{\mathbf{\Theta}_1,\dots,\mathbf{\Theta}_N \}$ denotes the scattering matrices of all PAs. By varying the PA positions, the segment lengths $\mathbf{x}$ are accordingly adjusted.

We consider both ideal PAs and DC-based in the optimization. For the ideal PAs, their scattering matrices can be fully reconfigured subject to the energy conservation constraint in \eqref{general_constraint}. The corresponding pinching beamforming optimization problem is given by 
\begin{subequations} \label{pinching_beamforming_pro_1}
    \begin{align}
        \max_{\mathbf{x}, \widetilde{\mathbf{\Theta}}} \quad & \mathcal{H}(\mathbf{x}, \widetilde{\mathbf{\Theta}}) \\
        \mathrm{s.t.} \quad & \mathbf{\Theta}_n^H \mathbf{\Theta}_n \preceq \mathbf{I}, \, \forall n=1,\dots,N, \\
        & x_n \ge \Delta x_{\min}, \,\forall n=1,\dots,N-1, \\
        & \sum_{n=0}^N x_n \le x_{\max}.
    \end{align}
\end{subequations}
In this problem, $\Delta x_{\min}$ denotes minimum spacing between PAs to avoid overlap and coupling, and $x_{\max}$ denotes the maximum length of the waveguide. The algorithm for solving this optimization problem is presented in Section \ref{appendix_algoritm_1}. 

For the DC-based PAs, the reconfiguration of the PA scattering matrix is realized by adjusting the coupling coefficient $\kappa$. The corresponding pinching beamforming optmization problem is given by  
\begin{subequations} \label{pinching_beamforming_pro_2}
    \begin{align}
        \max_{\mathbf{x}, \widetilde{\mathbf{\Theta}}} \quad & \mathcal{H}(\mathbf{x}, \widetilde{\mathbf{\Theta}}) \\
        \mathrm{s.t.} \quad & \mathbf{\Theta}_n = \begin{bmatrix}
        0 & \widetilde{\Theta}_1(\kappa_n) & \widetilde{\Theta}_2(\kappa_n) \\
        \widetilde{\Theta}_1(\kappa_n) & 0 & 0 \\
        \widetilde{\Theta}_2(\kappa_n) & 0 & 0
        \end{bmatrix},  \nonumber \\ 
        & \hspace{3.3cm} \forall n=1,\dots,N, \\
        & \kappa_n \in [0,1), \forall n = 1,\dots, N, \\
        & x_n \ge \Delta x_{\min}, \, \forall n=1,\dots,N-1, \\
        & \sum_{n=0}^N x_n \le x_{\max}.
    \end{align}
\end{subequations}
The algorithm for solving this optimization problem is presented in Section \ref{appendix_algoritm_2}. 

\begin{figure}[t!]
  \centering
  \includegraphics[width=0.48\textwidth]{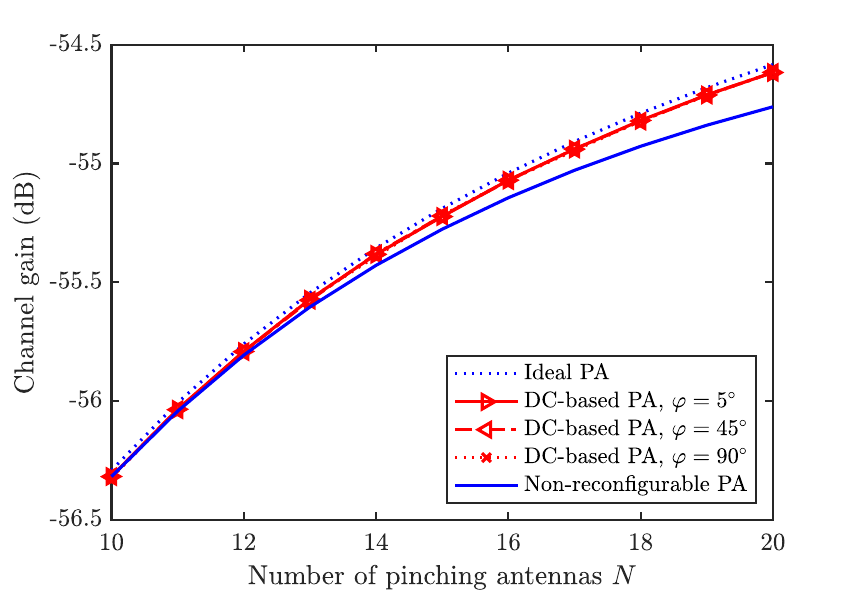}
  \caption{Channel gain versus the number of PAs when $\Delta x_{\min} = 0.5$ m.}
  \label{figure_sim_1}

  \includegraphics[width=0.48\textwidth]{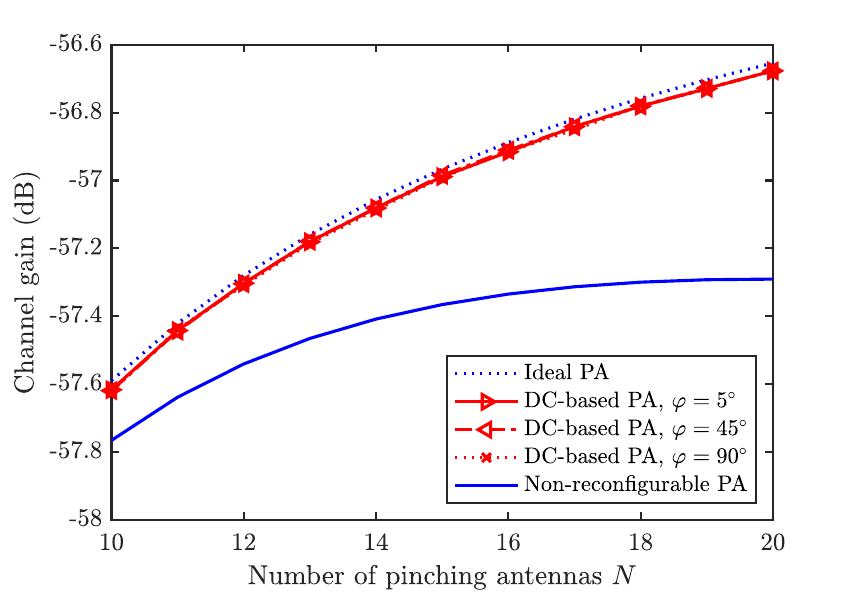}
  \caption{Channel gain versus the number of PAs when $\Delta x_{\min} = 1$ m.}
  \label{figure_sim_2}
\end{figure} 

\subsection{Numerical Results}
In this section, we present numerical simulation results to evaluate the proposed reconfigurable PASS. Unless otherwise specified, the following simulation setup is adopted. The signal frequency is set to $f = 15$ GHz, and the effective refractive index of the waveguide is $n_g = 1.4$. For the wireless channel parameters, we set $D_0 = 1$ m, $C_0 = -28$ dB, and $\alpha=1$. The waveguide is placed at $(y_g, z_g) = (0,3)$ m, while the receiver is located at $(15, 0, 0)$ m. The maximum waveguide length is set to $x_{\max} = 30$ m. All the results are generated using the pinching beamforming algorithms provided in Section \ref{sec:method}. To ensure near-optimality of the iterative algorithm for DC-based PAs, the best objective value among $100$ random initializations is selected for illustration.

For performance comparison, we adopt non-reconfigurable PAs as the baseline, which represents the most common model in the literature \cite{ding2025flexible, wang2025modeling}. In this case, PAs are assumed to have no capability for amplitude or phase control, and each PA radiates the same proportion of total power. Consequently, pinching beamforming can only be achieved by adjusting the signal phases through the placement of the PAs.

In Fig. \ref{figure_sim_1} and \ref{figure_sim_2}, we compare the achievable channel gains across different PA models as the number of PAs varies, for cases of $\Delta x_{\min} = 0.5$ m and $\Delta x_{\min} = 1$ m, respectively. The ideal PA attains the highest gain in both cases because it can fully control the per-PA amplitude and phase, whereas the non-reconfigurable model performs the worst. The DC-based PA is slightly inferior to the ideal case due to inherent coupling between its amplitude and phase controls.

The performance gap is small in Fig. \ref{figure_sim_1}, where $\Delta x_{\min}=0.5$ m. This is because in our considered single-user setting, amplitude control is the dominant form of reconfigurability, while phase reconfigurability is less critical because phase alignment can be achieved by repositioning the PAs. The required displacement of PA to sweep a full $2\pi$ phase is $\Delta d = 2\pi/\beta_g \approx 0.015$ m under our parameters, which has a negligible effect on path loss. Moreover, the short aperture at $\Delta x_{\min}=0.2$ m clusters the PAs, making per-PA pathloss nearly uniform and reducing the benefit of amplitude control. Consequently, all models exhibit similar performance, especially when the number of PAs is small. By contrast, with a larger spacing (i.e., $\Delta x_{\min}=1$ m), per-PA distances vary more significantly, producing larger path-loss disparities. In this case, amplitude control becomes crucial for “water-filling” power toward stronger links, leading to the larger performance gap observed in Fig. \ref{figure_sim_2}.

\begin{figure}[t!]
  \centering
  \includegraphics[width=0.48\textwidth]{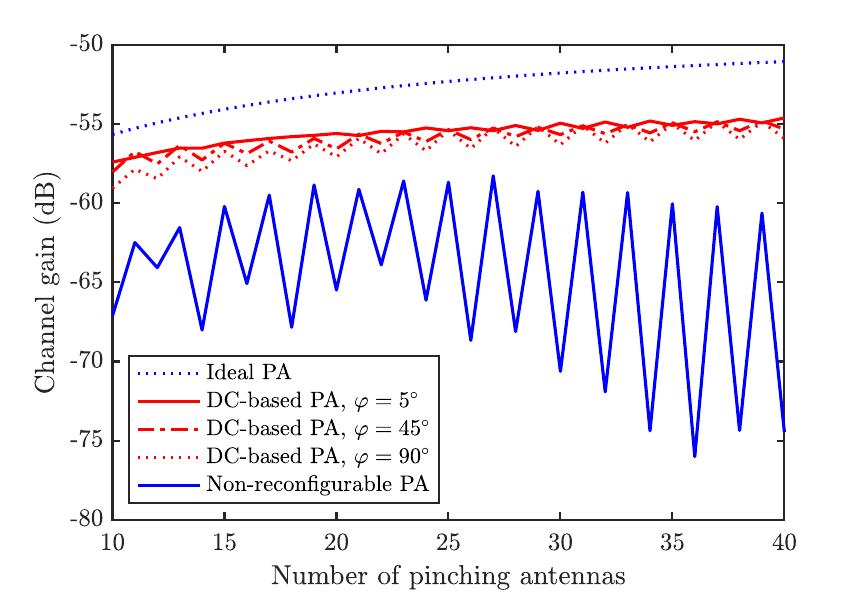}
  \caption{Channel gain versus the number of PAs with fixed positions and $\Delta x_{\min} = 0.2$ m}
  \label{figure_sim_3}
\end{figure} 

From the above discussion, achieving precise phase control by physically repositioning the PAs can be challenging in practice, since it requires millimeter-level adjustments, i.e., displacements well below $\Delta d \approx 0.015$ m, which drives hardware complexity and cost. Therefore, in Fig. \ref{figure_sim_3}, we consider a low-complexity setting with fixed PA positions under the simplest heuristic, i.e., placing the PAs at the minimum spacing $\Delta x_{\min}$ and as close to the receiver as possible to reduce pathloss. In this regime, the non-reconfigurable PA baseline exhibits highly unstable performance because it cannot correct phase misalignment at the fixed geometry. By contrast, the ideal PA maintains high gain since it controls phase without moving the PAs. The DC-based PA lies in between but shows a noticeable loss relative to the ideal case due to the intrinsic coupling between its amplitude and phase controls. Furthermore, the advantage of the wide achievable phase range of DC-based PAs (e.g., with $\varphi = 5^{\degree}$) is not significant. This is also due to the intrinsic coupling between amplitude and phase control, whereby achieving a larger phase shift necessarily reduces the signal amplitude, and vice versa (cf. Fig. \ref{Fig_ACPA_1}).

It is worth noting that the single-user setting considered here reduces the importance of phase reconfigurability and, in some cases, amplitude reconfigurability, so the performance gains over non-reconfigurable PAs can be modest. By contrast, in multi-user settings reconfigurable PAs are far more compelling, where amplitude control can allocate power both across links and across users, and stronger phase control is crucial for beam alignment and for mitigating inter-user interference.

\section{Discussion}
The multiport network model developed in this article provided a physically consistent foundation for interpreting PASS and for translating circuit-level constraints into end-to-end communication behavior. The established single-PA and multi-PA models clarified how load mismatch, internal reflections, and the wireless link jointly determine the received signal, while also revealing a matched special case that yields compact input-output models for analysis and design. Building on this general model, we presented two models for reconfigurable PAs.

For ideal, fully reconfigurable PAs, the best achievable gain is attained by aligning the aggregate PA phase with the channel and “water-filling” amplitudes across PA positions. This clarifies that, in single-user scenarios with optimized PA positions, amplitude control is the dominant degree of freedom, whereas fine phase control can often be realized geometrically by moving PAs over fractions of a wavelength. The DC-based PA yields a matched, reflection-free three-port when the even/odd mode impedances satisfy the product rule. This hardware structure offers a contactless, low-loss path to reconfigurability, but it inherently couples amplitude and phase. Very wide controllable phase ranges require extreme precision control of coupling coefficient and, therefore, more complex actuation. Although the DC-based architecture is theoretically matched and exhibit continuous control capability, its practical realizations must contend with tolerances in waveguide and coupler fabrication and finite-resolution in MEMS gap control. In particular, the residual mismatch at PA-waveguide connectors can reintroduce reflections, which would then be captured by the general multiport equations rather than the matched simplification. 

For future works, multi-user PASS is particularly compelling because amplitude control can allocate power across users while phase control suppresses inter-user interference. Finally, although extending the model to uplink is conceptually straightforward, a complete end-to-end uplink model and the associated design insights remain important topics for future work.

\section{Methods} \label{sec:method}

\subsection{Pinching Beamforming with Ideal PAs} \label{appendix_algoritm_1}
In this section, we elaborate on the proposed algorithm for solving the pinching beamforming optimization problem \eqref{pinching_beamforming_pro_1}. In the following, we first show that there exists an optimal reconfigurable scattering matrices $\widetilde{\mathbf{\Theta}}$ for PAs, and then optimizing the PA positions accordingly. 

\subsubsection{Optimal Scattering Matrices} To obtain the optimal $\widetilde{\mathbf{\Theta}}$, let us consider the following inequalities regarding the overall channel gain $\mathcal{H}(\mathbf{x}, \widetilde{\mathbf{\Theta}})$: 
\begin{align} \label{channel_gain_inequality}
    \mathcal{H}(\mathbf{x}, \widetilde{\mathbf{\Theta}}) = & \frac{ \big|  \mathbf{h}_{TR}^T \boldsymbol{\phi}_T \big|^2 }{1 + |\phi_R|^2 + 2 \Re \big\{\phi_R e^{-j2\beta x_0} \big\}} \nonumber \\
    \le & \frac{ \big\| \mathbf{h}_{TR} \big\|^2 \, \big\| \boldsymbol{\phi}_T \big\|^2  }{1 - | \phi_R |^2}.
\end{align}
The upper bound in this inequality achieves when $\boldsymbol{\phi}_T$ and $\phi_R$ take the form:
\begin{align} \label{optimal_structure}
    \boldsymbol{\phi}_T^{\star} = \theta_T \frac{\mathbf{h}_{TR}^*}{\|\mathbf{h}_{TR}\|}, \quad \phi_{R}^{\star} = \theta_R e^{j 2\beta x_0 - j \frac{\pi}{2}},
\end{align}
where $\theta_T \ge 0$ and $\theta_R \ge 0$ are the non-negative real-valued amplitude factors. Substituting \eqref{optimal_structure} into \eqref{channel_gain_inequality}, we obtain
\begin{equation} \label{channel_gain_inequality_2}
    \mathcal{H}(\mathbf{x}, \widetilde{\mathbf{\Theta}}) \le \frac{\theta_T^2 \big\| \mathbf{h}_{TR} \big\|^2}{1 - \theta_R^2}.
\end{equation}

Furthermore, following the energy conservation conditions, it must hold that 
\begin{equation}
    \|\boldsymbol{\phi}_T^{\star}\|^2 + |\phi_{R}^{\star}|^2 \le 1 \quad \Rightarrow \quad \theta_T^2 + \theta_R^2 \le 1,
\end{equation}
which reflects that the total power radiated and reflected cannot exceed the input power. Then, it can be readily proved that the upper bound in \eqref{channel_gain_inequality_2} is maximized when $\theta_T^2 + \theta_R^2 = 1$, yielding
\begin{equation} \label{optimal_channel_gain}
    \mathcal{H}(\mathbf{x}, \widetilde{\mathbf{\Theta}})  \le \frac{\theta_T^2 \big\| \mathbf{h}_{TR} \big\|^2}{1 - \theta_R^2} =  \big\| \mathbf{h}_{TR} \big\|^2,
\end{equation}
which is independent of the specific scattering matrices. Hence, for any given $\mathbf{x}$, the maximum channel gain is $\widetilde{\mathbf{\Theta}}$ is $\big\| \mathbf{h}_{TR} \big\|^2$, achieved whenever \eqref{optimal_structure} holds with $\theta_T^2 + \theta_R^2 = 1$. Without loss of optimality, setting $\theta_R^2 = 0$ gives the final optimal solution:
\begin{equation}
    \boldsymbol{\phi}_T^{\star} = \frac{\mathbf{h}_{TR}^*}{\|\mathbf{h}_{TR}\|}, \quad \phi_{R}^{\star} = 0.
\end{equation}

In the following, we demonstrate that the above optimal solution can be achieved by the matched scattering matrices of PAs, which takes the form:
\begin{equation}
    \mathbf{\Theta}_n = \begin{bmatrix}
        0 & \hat{\Theta}_{1,n} & \hat{\Theta}_{2,n} \\
        \hat{\Theta}_{1,n} & 0 & 0 \\
        \hat{\Theta}_{2,n} & 0 & 0
    \end{bmatrix}.
\end{equation}
Although this matrix exhibits a similar form as the matrix \eqref{DC_S_matrix} for DC-based PAs, the key difference here is that the scattering parameters $\hat{\Theta}_{1,n}$ and $\hat{\Theta}_{2,n}$ can be arbitrarily adjusted subject only to the energy conservation constraint:
\begin{equation} \label{energy_conservation_optimal}
    \big| \hat{\Theta}_{1,n} \big|^2 + \big| \hat{\Theta}_{2,n} \big|^2 \le 1.
\end{equation}    
Under this circumstance, the overall channel gain reduces to a form similar to \eqref{DC_overall_channel}, given by 
\begin{equation}
    \mathcal{H}(\mathbf{x}, \widetilde{\mathbf{\Theta}}) = \left| \sum_{n=1}^N h_{TR,n} \left( \hat{\Theta}_{2,n} \prod_{i=1}^{n-1} \hat{\Theta}_{1,i} \right) e^{-j\beta_g \sum_{i=0}^{n-1} x_i} \right|^2.
\end{equation}
Then, the optimal channel gain given in \eqref{optimal_channel_gain} can be achieved through the following two steps.
\begin{itemize}
    \item \textbf{Phase Alignment:} In this step, the phases of $\hat{\Theta}_{1,n}$ and $\hat{\Theta}_{2,n}$ are chosen such that the aggregate signal phase aligns with the wireless channel phase. Writing $h_{TR,n} = |h_{TR,n}|e^{-j\vartheta_n}$, the alignment condition is
    \begin{equation} \label{optimal_phase}
        \angle \hat{\Theta}_{2,n} + \sum_{i=1}^{n-1} \angle \hat{\Theta}_{1,i} - \beta_g \sum_{i=0}^{n-1}x_i = \theta_n \,\, (\mathrm{mod}\,\,2\pi).
    \end{equation}
    Since the phases $\angle \hat{\Theta}_{1,n}$ and $\angle \hat{\Theta}_{2,n}$ can be arbitrarily tuned for ideal PAs, infinitely many solutions exist. A simple choice is to fix $\angle \hat{\Theta}_{1,n} = 0$ for all $n$, in which case $\angle \hat{\Theta}_{2,n}$ can be directly determined.
    \item \textbf{Amplitude Alignment:} Once phases are aligned, the overall channel gain simplifies to  
    \begin{equation}
        \mathcal{H}(\mathbf{x}, \widetilde{\mathbf{\Theta}}) = \left| \sum_{n=1}^N \big| h_{TR,n} \big| \left( \big|\hat{\Theta}_{2,n}\big| \prod_{i=1}^{n-1} \big|\hat{\Theta}_{1,i}\big| \right) \right|^2.
    \end{equation} 
    To achieve the optimum value in \eqref{optimal_channel_gain}, the amplitudes must satisfy
    \begin{equation}
        \big|\hat{\Theta}_{2,n}\big| \prod_{i=1}^{n-1} \big|\hat{\Theta}_{1,i}\big| = \frac{|h_{TR,n}|}{\|\mathbf{h}_{TR}\|}, \, \forall n=1,\dots,N.
    \end{equation}
    One feasible solution subject to the energy conservation constraint \eqref{energy_conservation_optimal} is given by 
    \begin{align} \label{optimal_amp}
        \big|\hat{\Theta}_{1,n}\big|^2 = &\frac{\sum_{i=n+1}^N \big| h_{TR,i} \big|^2 }{\sum_{i=n}^N \big| h_{TR,i} \big|^2} , \nonumber \\
        \big|\hat{\Theta}_{2,n}\big|^2 = &\frac{ \big| h_{TR,n} \big|^2 }{\sum_{i=n}^N \big| h_{TR,i} \big|^2}.
    \end{align}
\end{itemize}

\subsubsection{Optimal PA Positions}

Based on the previous analysis, for any given $\mathbf{x}$, the optimal reconfigurable scattering matrices can be determined using \eqref{optimal_phase} and \eqref{optimal_amp}, denoted as $\widetilde{\mathbf{\Theta}}^{\star}(\mathbf{x})$. Substituting this result, the optimization problem \eqref{pinching_beamforming_pro_1} with respect to $\mathbf{x}$ can be reformulated as
\begin{subequations}
    \begin{align}
        \max_{\mathbf{x}} \quad & \mathcal{H}(\mathbf{x}, \widetilde{\mathbf{\Theta}}^{\star}(\mathbf{x})) = \sum_{n=1}^N \big| h_{TR,n} \big|^2 \\
        \label{constraint_1}
        \mathrm{s.t.} \quad & x_n \ge \Delta x_{\min}, \,\forall n=1,\dots,N-1, \\
        \label{constraint_2}
        & \sum_{n=0}^N x_n \le x_{\max},
    \end{align}
\end{subequations}
where the objective is obtained from \eqref{optimal_channel_gain}, and the channel coefficient $h_{TR,n}$ is defined in \eqref{wireless_channel_coefficients}. For convenience, let $s_n = \sum_{i=0}^{n-1} x_i$ denote the $x$-coordinate of the $n$-th PA. In terms of $\{s_n\}$, constraints \eqref{constraint_1}-\eqref{constraint_2} are equivalently expressed as
\begin{align}
&0 \le s_1 \le s_2 \le \cdots \le s_N \le x_{\max}, \\
&s_n - s_{n-1} \ge \Delta x_{\min}, \, \forall n=2,\dots,N.
\end{align}
Accordingly, the distance from the $n$-th PA to the receiver can be rewritten as
\begin{equation}
    d_n = \sqrt{\left(s_n - x_r\right)^2 + \xi  }.
\end{equation}  
Hence, optimizing over $\{s_n\}$ is equivalent to optimizing the original variables $\{x_n\}$. From \eqref{wireless_channel_coefficients}, the individual channel gain is given by 
\begin{equation}
    \big| h_{TR,n} \big|^2 = \left( \frac{\lambda}{4 \pi d_n} \right)^2,
\end{equation}
indicating that the optimization of PA positions reduces to minimizing the pathloss.
%
%
The new objective function decreases monotonically with $d_n$ and therefore also $|s_n-x_r|$. This implies that, at optimum, the inter-PA spacing must be tight:
\begin{equation} \label{PA_tight_condition}
    s_n - s_{n-1} = \Delta x_{\min}, \forall n=2,\dots,N.
\end{equation}
Therefore, the $N$ PAs form a rigid block of length $L_P = (N-1)\Delta x_{\min}$.  To maximize the overall channel gain, the block center, denoted by $\delta$, should be positioned as close to the receiver as possible while ensuring the PA block confined within the waveguide range. This yields the optimal center position as follows:
\begin{equation}
    \delta^\star = \min \left\{ \max \left\{x_r, \frac{L_P}{2}\right\}, x_{\max} - \frac{L_P}{2}  \right\}.
\end{equation}
Combining with the condition \eqref{PA_tight_condition}, the optimal PA positions are given by 
\begin{equation}
    s^{\star}_n = \delta^\star + \left( n - \frac{N+1}{2} \right) \Delta x_{\min}.
\end{equation}

\subsection{Pinching Beamforming with DC-based PAs} \label{appendix_algoritm_2}

In this section, we develop an algorithm for solving problem \eqref{pinching_beamforming_pro_2}. Unlike the ideal-PA case, this problem involves intrinsic coupling between amplitude and phase control, which makes finding the global optimum challenging. As a remedy, we adopt an alternating optimization approach, where the scattering matrices and PA positions are updated alternatingly.

We first consider the optimization of $\widetilde{\mathbf{\Theta}}$, i.e., the set of coupling coefficients $\{\kappa_n\}$, while keeping the PA positions $\mathbf{x}$ fixed. To solve this problem, we reparameterize the coupling coefficients through the coupled port amplitude \eqref{coupled_port_amplitude}:
\begin{equation}
    \kappa_n = \big|\tanh(\psi_n)\big|
\end{equation}
which automatically ensures the constraint $\kappa_n \in [0,1)$ for arbitrary $\psi_n$. With this transformation, the subproblem becomes unconstrained in terms of $\{\psi_n\}$ and can be efficiently solved using a quasi-Newton method \cite{nocedal2006numerical}. In particular, we adopt the Broyden-Fletcher-Goldfarb-Shanno (BFGS) update to approximate the Hessian matrix in the quasi-Newton method, resulting in a per-iteration computational complexity of $\mathcal{O}(N^2)$~\cite{nocedal2006numerical}.

Furthermore, for fixed $\widetilde{\mathbf{\Theta}}$, the subproblem with respect to $\mathbf{x}$, or equivalently $\{s_n\}$, can be solved through element-wise optimization, where the positions $\{s_n\}$ are optimized sequentially using coordinate ascent. In particular, since the valuables $\{s_n\}$ do not coupled, each $s_n$ can be optimized individually via a one-dimensional grid search of the overall channel gain over the following feasible interval:
\begin{equation}
    s_n = [ s_{n-1} + \Delta x_{\min}, s_{n+1} - \Delta x_{\min} ]\, \cap \, [0, x_{\max}].
\end{equation}
Let $Q$ denote the number of grid points used in the one-dimensional search. The computational complexity of sweeping all $\{s_n\}_{n=1}^N$ in each iteration is thus $\mathcal{O}(NQ)$.   

Based on the optimization methods for both the scattering matrices and the PA positions, these two blocks of variables can be optimized in an alternating manner until the overall channel gain improvement falls below a predefined tolerance. 

\section{Data Availability}
All data supporting this study are generated from the model equations and algorithms described in the paper and can be reproduced in MATLAB. No third-party datasets were used or analyzed.

\section{Code Availability}
The underlying code for this study is not publicly available but may be made available to qualified researchers on reasonable request from the corresponding author.

\section{Author Contributions}
Z.W. and J.X. conceived this idea and conducted the theoretical analysis. J.X. developed the initial model using multiport network theory. Z.W. refined this model, derived the end-to-end signal representation, proposed two reconfigurable models, and developed the beamforming optimization algorithms. C.O., X.M., and Y.L. streamlined the overall article structure and improved the presentation. Y.L. directed and supervised the research. All authors discussed the results and reviewed and edited the manuscript.

\section{Competing Interest}
The authors declare no competing interests.

\balance
\bibliographystyle{naturemag}
\bibliography{mybib}

\begin{thebibliography}{10}
\expandafter\ifx\csname url\endcsname\relax
  \def\url#1{\texttt{#1}}\fi
\expandafter\ifx\csname urlprefix\endcsname\relax\def\urlprefix{URL }\fi
\providecommand{\bibinfo}[2]{#2}
\providecommand{\eprint}[2][]{\url{#2}}

\bibitem{saad2019vision}
\bibinfo{author}{Saad, W.}, \bibinfo{author}{Bennis, M.} \& \bibinfo{author}{Chen, M.}
\newblock \bibinfo{title}{A vision of 6{G} wireless systems: Applications, trends, technologies, and open research problems}.
\newblock \emph{\bibinfo{journal}{{IEEE} Netw.}} \textbf{\bibinfo{volume}{34}}, \bibinfo{pages}{134--142} (\bibinfo{year}{2020}).

\bibitem{larsson2014massive}
\bibinfo{author}{Larsson, E.~G.}, \bibinfo{author}{Edfors, O.}, \bibinfo{author}{Tufvesson, F.} \& \bibinfo{author}{Marzetta, T.~L.}
\newblock \bibinfo{title}{Massive {MIMO} for next generation wireless systems}.
\newblock \emph{\bibinfo{journal}{{IEEE} Commun. Mag.}} \textbf{\bibinfo{volume}{52}}, \bibinfo{pages}{186--195} (\bibinfo{year}{2014}).

\bibitem{wong2020fluid}
\bibinfo{author}{Wong, K.-K.}, \bibinfo{author}{Shojaeifard, A.}, \bibinfo{author}{Tong, K.-F.} \& \bibinfo{author}{Zhang, Y.}
\newblock \bibinfo{title}{Fluid antenna systems}.
\newblock \emph{\bibinfo{journal}{{IEEE} Trans. Wireless Commun.}} \textbf{\bibinfo{volume}{20}}, \bibinfo{pages}{1950--1962} (\bibinfo{year}{2020}).

\bibitem{new2024tutorial}
\bibinfo{author}{New, W.~K.} \emph{et~al.}
\newblock \bibinfo{title}{A tutorial on fluid antenna system for 6{G} networks: Encompassing communication theory, optimization methods and hardware designs}.
\newblock \emph{\bibinfo{journal}{{IEEE} Commun. Surv. Tut.}} \textbf{\bibinfo{volume}{27}}, \bibinfo{pages}{2325--2377} (\bibinfo{year}{2025}).

\bibitem{zhu2023movable}
\bibinfo{author}{Zhu, L.}, \bibinfo{author}{Ma, W.} \& \bibinfo{author}{Zhang, R.}
\newblock \bibinfo{title}{Movable antennas for wireless communication: Opportunities and challenges}.
\newblock \emph{\bibinfo{journal}{{IEEE} Commun. Mag.}} \textbf{\bibinfo{volume}{62}}, \bibinfo{pages}{114--120} (\bibinfo{year}{2024}).

\bibitem{ning2025movable}
\bibinfo{author}{Ning, B.} \emph{et~al.}
\newblock \bibinfo{title}{Movable antenna-enhanced wireless communications: General architectures and implementation methods}.
\newblock \emph{\bibinfo{journal}{{IEEE} Wireless Commun.}} \textbf{\bibinfo{volume}{early access}} (\bibinfo{year}{2025}).

\bibitem{bai2022dynamically}
\bibinfo{author}{Bai, Y.} \emph{et~al.}
\newblock \bibinfo{title}{A dynamically reprogrammable surface with self-evolving shape morphing}.
\newblock \emph{\bibinfo{journal}{Nature}} \textbf{\bibinfo{volume}{609}}, \bibinfo{pages}{701--708} (\bibinfo{year}{2022}).

\bibitem{an2025flexible}
\bibinfo{author}{An, J.} \emph{et~al.}
\newblock \bibinfo{title}{Flexible intelligent metasurfaces for downlink multiuser {MISO} communications}.
\newblock \emph{\bibinfo{journal}{{IEEE} Trans. Wireless Commun.}} \textbf{\bibinfo{volume}{early access}} (\bibinfo{year}{2025}).

\bibitem{suzuki2022pinching}
\bibinfo{author}{Fukuda, A.}, \bibinfo{author}{Yamamoto, H.}, \bibinfo{author}{Okazaki, H.}, \bibinfo{author}{Suzuki, Y.} \& \bibinfo{author}{Kawai, K.}
\newblock \bibinfo{title}{Pinching antenna: Using a dielectric waveguide as an antenna}.
\newblock \emph{\bibinfo{journal}{NTT DOCOMO Technical J.}} \textbf{\bibinfo{volume}{23}}, \bibinfo{pages}{5--12} (\bibinfo{year}{2022}).

\bibitem{liu2025pinching}
\bibinfo{author}{Liu, Y.} \emph{et~al.}
\newblock \bibinfo{title}{Pinching-antenna systems ({PASS}): Architecture designs, opportunities, and outlook}.
\newblock \emph{\bibinfo{journal}{{IEEE} Commun. Mag.}} \textbf{\bibinfo{volume}{accepted to appear}} (\bibinfo{year}{2025}).

\bibitem{ding2025flexible}
\bibinfo{author}{Ding, Z.}, \bibinfo{author}{Schober, R.} \& \bibinfo{author}{Poor, H.~V.}
\newblock \bibinfo{title}{Flexible-antenna systems: A pinching-antenna perspective}.
\newblock \emph{\bibinfo{journal}{{IEEE} Trans. Commun.}} \textbf{\bibinfo{volume}{early access}} (\bibinfo{year}{2025}).

\bibitem{wang2025modeling}
\bibinfo{author}{Wang, Z.}, \bibinfo{author}{Ouyang, C.}, \bibinfo{author}{Mu, X.}, \bibinfo{author}{Liu, Y.} \& \bibinfo{author}{Ding, Z.}
\newblock \bibinfo{title}{Modeling and beamforming optimization for pinching-antenna systems}.
\newblock \emph{\bibinfo{journal}{arXiv preprint arXiv:2502.05917}}  (\bibinfo{year}{2025}).

\bibitem{tegos2025minimum}
\bibinfo{author}{Tegos, S.~A.}, \bibinfo{author}{Diamantoulakis, P.~D.}, \bibinfo{author}{Ding, Z.} \& \bibinfo{author}{Karagiannidis, G.~K.}
\newblock \bibinfo{title}{Minimum data rate maximization for uplink pinching-antenna systems}.
\newblock \emph{\bibinfo{journal}{{IEEE} Wireless Commun. Lett.}} \textbf{\bibinfo{volume}{14}}, \bibinfo{pages}{1516--1520} (\bibinfo{year}{2025}).

\bibitem{xiao2025frequency}
\bibinfo{author}{Xiao, J.}, \bibinfo{author}{Wang, J.}, \bibinfo{author}{Zeng, M.}, \bibinfo{author}{Liu, Y.} \& \bibinfo{author}{Karagiannidis, G.~K.}
\newblock \bibinfo{title}{Frequency-selective modeling and analysis for {OFDM}-integrated wideband pinching-antenna systems}.
\newblock \emph{\bibinfo{journal}{{IEEE} Wireless Commun. Lett.}} \textbf{\bibinfo{volume}{early access}} (\bibinfo{year}{2025}).

\bibitem{chen2025dynamic}
\bibinfo{author}{Chen, J.-C.}, \bibinfo{author}{Wu, P.-C.} \& \bibinfo{author}{Wong, K.-K.}
\newblock \bibinfo{title}{Dynamic placement of pinching antennas for multicast {MU-MISO} downlinks}.
\newblock \emph{\bibinfo{journal}{IEEE Open J. Commun. Soc.}} \textbf{\bibinfo{volume}{6}}, \bibinfo{pages}{5611--5625} (\bibinfo{year}{2025}).

\bibitem{ouyang2025array}
\bibinfo{author}{Ouyang, C.}, \bibinfo{author}{Wang, Z.}, \bibinfo{author}{Liu, Y.} \& \bibinfo{author}{Ding, Z.}
\newblock \bibinfo{title}{Array gain for pinching-antenna systems ({PASS})}.
\newblock \emph{\bibinfo{journal}{{IEEE} Commun. Lett.}} \textbf{\bibinfo{volume}{29}}, \bibinfo{pages}{1471--1475} (\bibinfo{year}{2025}).

\bibitem{jiang2025pinching}
\bibinfo{author}{Jiang, H.}, \bibinfo{author}{Wang, Z.} \& \bibinfo{author}{Liu, Y.}
\newblock \bibinfo{title}{Pinching-antenna system ({PASS}) enhanced covert communications}.
\newblock \emph{\bibinfo{journal}{arXiv preprint arXiv:2504.10442}}  (\bibinfo{year}{2025}).

\bibitem{papanikolaou2025resolving}
\bibinfo{author}{Papanikolaou, V.~K.} \emph{et~al.}
\newblock \bibinfo{title}{Resolving the double near-far problem via wireless powered pinching-antenna networks}.
\newblock \emph{\bibinfo{journal}{{IEEE} Wireless Commun. Lett.}} \textbf{\bibinfo{volume}{early access}} (\bibinfo{year}{2025}).

\bibitem{xu2025pinching}
\bibinfo{author}{Xu, X.}, \bibinfo{author}{Mu, X.}, \bibinfo{author}{Wang, Z.}, \bibinfo{author}{Liu, Y.} \& \bibinfo{author}{Nallanathan, A.}
\newblock \bibinfo{title}{Pinching-antenna systems ({PASS}): Power radiation model and optimal beamforming design}.
\newblock \emph{\bibinfo{journal}{arXiv preprint arXiv:2505.00218}}  (\bibinfo{year}{2025}).

\bibitem{xu2025joint}
\bibinfo{author}{Xu, Y.} \emph{et~al.}
\newblock \bibinfo{title}{Joint radiation power, antenna position, and beamforming optimization for pinching-antenna systems with motion power consumption}.
\newblock \emph{\bibinfo{journal}{arXiv preprint arXiv:2507.02348}}  (\bibinfo{year}{2025}).

\bibitem{6880934}
\bibinfo{author}{Ivrlač, M.~T.} \& \bibinfo{author}{Nossek, J.~A.}
\newblock \bibinfo{title}{The multiport communication theory}.
\newblock \emph{\bibinfo{journal}{IEEE Circuits Syst. Mag.}} \textbf{\bibinfo{volume}{14}}, \bibinfo{pages}{27--44} (\bibinfo{year}{2014}).

\bibitem{10373407}
\bibinfo{author}{Mezghani, A.} \emph{et~al.}
\newblock \bibinfo{title}{Reincorporating circuit theory into information theory}.
\newblock \emph{\bibinfo{journal}{IEEE BITS Inf. Theory Mag.}} \textbf{\bibinfo{volume}{4}}, \bibinfo{pages}{40--58} (\bibinfo{year}{2024}).

\bibitem{ivrlavc2010toward}
\bibinfo{author}{Ivrla{\v{c}}, M.~T.} \& \bibinfo{author}{Nossek, J.~A.}
\newblock \bibinfo{title}{Toward a circuit theory of communication}.
\newblock \emph{\bibinfo{journal}{IEEE Trans. Circuits Syst. I: Regul. Pap.}} \textbf{\bibinfo{volume}{57}}, \bibinfo{pages}{1663--1683} (\bibinfo{year}{2010}).

\bibitem{11006094}
\bibinfo{author}{Pizzo, A.} \& \bibinfo{author}{Lozano, A.}
\newblock \bibinfo{title}{Mutual coupling in holographic {MIMO}: Physical modeling and information-theoretic analysis}.
\newblock \emph{\bibinfo{journal}{IEEE J. Sel. Areas Inf. Theory}} \textbf{\bibinfo{volume}{6}}, \bibinfo{pages}{111--126} (\bibinfo{year}{2025}).

\bibitem{shen2021modeling}
\bibinfo{author}{Shen, S.}, \bibinfo{author}{Clerckx, B.} \& \bibinfo{author}{Murch, R.}
\newblock \bibinfo{title}{Modeling and architecture design of reconfigurable intelligent surfaces using scattering parameter network analysis}.
\newblock \emph{\bibinfo{journal}{{IEEE} Trans. Wireless Commun.}} \textbf{\bibinfo{volume}{21}}, \bibinfo{pages}{1229--1243} (\bibinfo{year}{2021}).

\bibitem{11098513}
\bibinfo{author}{Xu, J.}, \bibinfo{author}{Wang, H.}, \bibinfo{author}{Liu, R.}, \bibinfo{author}{Nossek, J.~A.} \& \bibinfo{author}{Swindlehurst, A.~L.}
\newblock \bibinfo{title}{Non-reciprocal reconfigurable intelligent surfaces}.
\newblock \emph{\bibinfo{journal}{{IEEE} Wireless Commun. Lett.}} \textbf{\bibinfo{volume}{early access}} (\bibinfo{year}{2025}).

\bibitem{9048753}
\bibinfo{author}{Marzetta, T.~L.}
\newblock \bibinfo{title}{Super-directive antenna arrays: Fundamentals and new perspectives}.
\newblock In \emph{\bibinfo{booktitle}{Asilomar Conf. Signals, Syst., Comput. (ACSSC)}}, \bibinfo{pages}{1--4} (\bibinfo{year}{2019}).

\bibitem{mongia1999rf}
\bibinfo{author}{Mongia, R.}, \bibinfo{author}{Bahl, I.~J.} \& \bibinfo{author}{Bhartia, P.}
\newblock \emph{\bibinfo{title}{RF and microwave coupled-line circuits}} (\bibinfo{publisher}{Norwood, MA: Artech House}, \bibinfo{year}{1999}).

\bibitem{pozar2021microwave}
\bibinfo{author}{Pozar, D.~M.}
\newblock \emph{\bibinfo{title}{Microwave Engineering}} (\bibinfo{publisher}{Hoboken, NJ, USA: Wiley}, \bibinfo{year}{2012}).

\bibitem{tse2005fundamentals}
\bibinfo{author}{Tse, D.} \& \bibinfo{author}{Viswanath, P.}
\newblock \emph{\bibinfo{title}{Fundamentals of wireless communication}} (\bibinfo{publisher}{Cambridge, U.K.: Cambridge Univ. Press}, \bibinfo{year}{2005}).

\bibitem{6568978}
\bibinfo{author}{Shah, U.}, \bibinfo{author}{Sterner, M.} \& \bibinfo{author}{Oberhammer, J.}
\newblock \bibinfo{title}{High-directivity {MEMS}-tunable directional couplers for 10-18-{GHz} broadband applications}.
\newblock \emph{\bibinfo{journal}{IEEE Trans. Microw. Theory Tech.}} \textbf{\bibinfo{volume}{61}}, \bibinfo{pages}{3236--3246} (\bibinfo{year}{2013}).

\bibitem{nocedal2006numerical}
\bibinfo{author}{Nocedal, J.} \& \bibinfo{author}{Wright, S.}
\newblock \emph{\bibinfo{title}{Numerical optimization}} (\bibinfo{publisher}{New York, NY, USA: Springer-Verlag}, \bibinfo{year}{2006}).

\end{thebibliography}

\end{document}